\documentstyle[psfig]{mn}
\newcommand{\ms}{\rm M_\odot}
\newcommand{\ls}{\rm L_\odot}
\newcommand{\lfir}{L_{\rm FIR}}
\newcommand{\lbol}{L_{\rm bol}}
\newcommand{\lb}{L_{\rm B}}

\newcommand{\urad}{U_{\rm rad}}
\newcommand{\ub}{U_{B}}

\newcommand{\hii}{H\,{\sc ii}\ }
\newcommand{\hi}{H\,{\sc i}\ }
\newcommand{\anth}{\alpha_{nth}}

\newcommand{\prad}{P_{\rm rad}}
\newcommand{\pnu}{P_{1.49GHz}}

\newcommand{\apjs}{ApJS, }
\newcommand{\apj}{ApJ, }
\newcommand{\aajou}{A\&A, }
\newcommand{\aas}{A\&AS, }
\newcommand{\mnras}{MNRAS, }

\begin{document}

\title[Cosmic ray propagation and star formation history of NGC~1961]
{Cosmic ray propagation and star formation history of NGC~1961}
\author[U.~Lisenfeld et al.]
{U.~Lisenfeld$^{1,2}$
\thanks{Present address:
IRAM, Avenida Divina Pastora 7, N\'ucleo Central, 
18012 Granada, Spain, e-mail ute@iram.es},
P.~Alexander $^1$, G.G.~Pooley$^1$ and T.~Wilding$^1$\\
$^1$ Mullard Radio Astronomy Observatory,
        Cavendish Laboratory,
        Madingley Road,
        Cambridge CB3 OHE, UK \\
$^2$ Universidad de Granada,
Facultad de F\'{\i}sica Te\'orica y del Cosmos,
18002 Granada, Spain}
\maketitle
\begin{abstract}
We present new radio continuum data at 4 frequencies on the
supermassive, peculiar galaxy NGC~1961. These observations
allow us to separate the thermal and the nonthermal radio emission
and to determine the nonthermal spectral index distribution.
This spectral index distribution in
the galactic disk is unusual: at the
maxima of the radio emission the synchrotron spectrum is very
steep, indicating aged cosmic ray electrons. 
Away from the maxima the spectrum is much flatter.
The steep spectrum of the synchrotron emission at the maxima indicates
that a strong decline of the star formation rate 
has taken place at these sites. 
The extended radio emission is a sign of 
recent cosmic ray acceleration, probably  by   recent star
formation. 
We suggest that  
a violent event in the past, most likely a merger or a
collision with an  intergalactic gas cloud, 
has caused the various unusual
features of the galaxy. 
\end{abstract}


\section{Introduction}

NGC~1961 (Arp 184) is a very massive (dynamical mass $10^{12} - 10^{13}\ms$,
Gottesman et al. 1983) Sb spiral galaxy with a peculiar and asymmetric optical
appearance.
It is among the  most massive and largest known spiral galaxies
(Romanishin 1983); its  optical diameter of $D_{25}=4.6$ arcmin 
(de Vaucouleurs et al. 1991, RC3) translates to 110 kpc 
(adopting a distance of 82.9 Mpc for $H_0 = 50$ km s$^{-1}$ Mpc$^{-1}$).
It is a member of a group of 6 galaxies
where it is by far the largest and most massive member.

Despite its size and peculiar appearance, NGC~1961 has received relatively little attention.
A detailed study of its \hi structure and kinematics by Shostak et
al. (1982, SHSVV)
found an unusual asymmetric morphology with an extensive ``wing''
of gas extending 30 kpc to the north-west and a sharp edge to the south-east.
SHSVV suggested that the 
disturbed appearance is due to stripping of the gas in NGC~1961
by a hot intergalactic medium (IGM). They discarded the possibility
of tidal interaction with another galaxy since there is 
no galaxy with a large enough mass sufficiently close to NGC~1961. 
They also dismissed the possibility
of a merger with another galaxy or an intergalactic gas cloud since they
found no trace of a merger remnant such as a second nucleus, or disturbed 
velocity distribution in the \hi gas.
In favour of the stripping hypothesis SHSVV noted the following points.
i) The displacement of the \hi with respect
to the optical disk in the south of the galaxy.
ii) The presence of increased
radio continuum emission in the south-east which they interpreted as being
caused by an enhancement of the magnetic field or by an increase
in the cosmic ray
(CR) density due to a high star formation rate (SFR).
iii) The
blue colour of the optical emission  of the NW wing and 
the south-eastern ridge indicating
recent star formation.
iv) A curved  X-ray emission feature in the south-east of the 
galaxy detected by the {\it Einstein} satellite
(Shostak et al. 1982) which they interpreted as emission from hot
IGM. However, recent ROSAT observations have shown that
the X-ray emission of NGC~1961 is fairly typical for a spiral galaxy
and comes mainly from the central region (Pence \& Rots 1997). 
The cause of the 
asymmetric \hi distribution therefore remains unclear.

In this paper we present multi-frequency radio continuum data which
we use to identify thermal and nonthermal (synchrotron) emission. 
The thermal radio emission reveals the sites and intensity of 
present day star formation, 
whereas the synchroton emission, and in particular
its spectral shape, contains information about the
star formation in the past (on time-scales of a few $10^7$ yr which
is related to the life-time of the CR electrons
and of supernova (SN) progenitors), and also  
about the acceleration and propagation mechanisms of CRs.

\section{Observations}

\begin{figure*}
\begin{minipage}{178mm}
\centerline{\hskip-0.5cm 1.46~GHz \hskip7.5cm    4.92~GHz} 
\hbox{
 \psfig{file=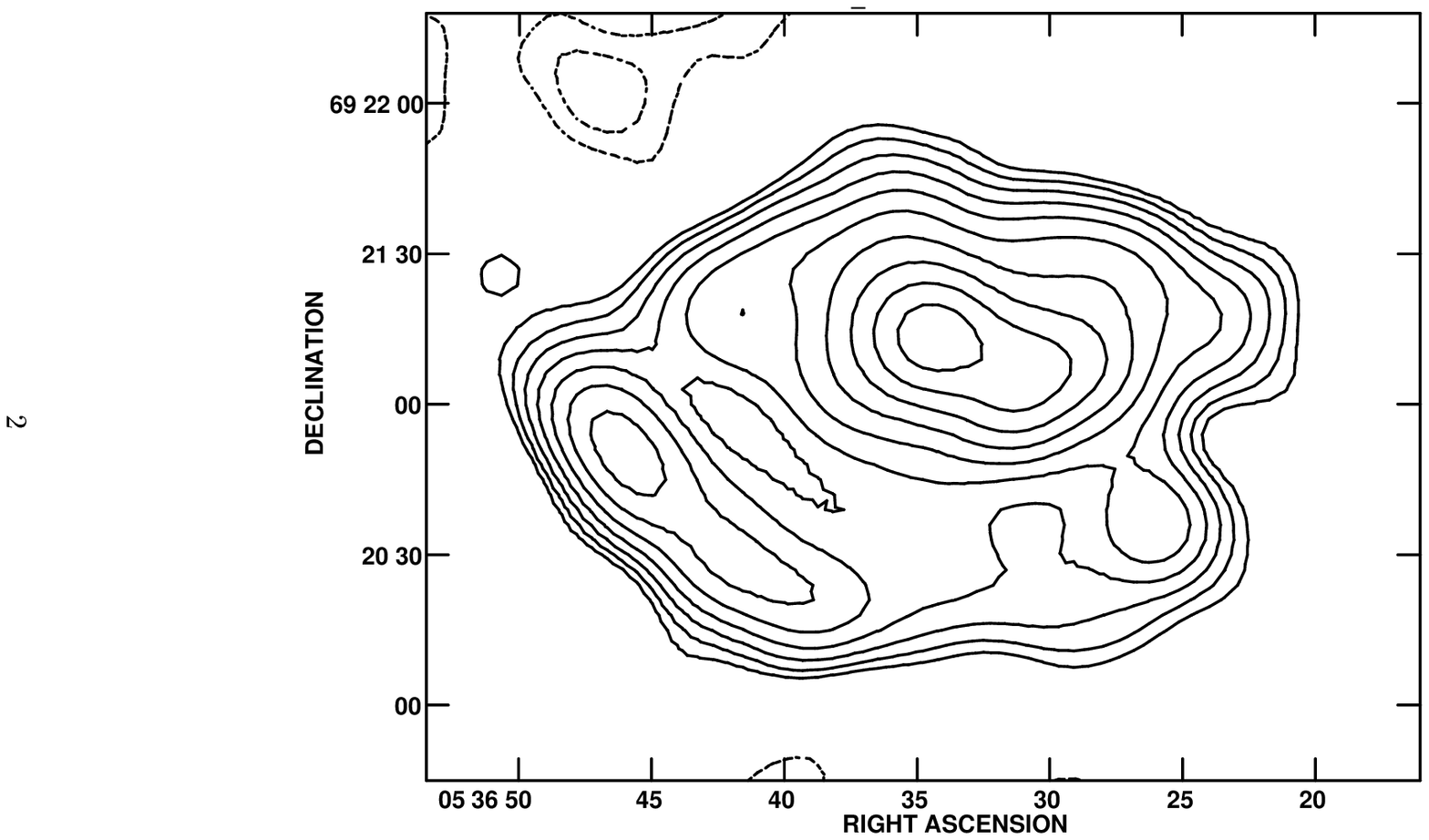,width=8.7cm,clip=,angle=270}\quad
 \psfig{file=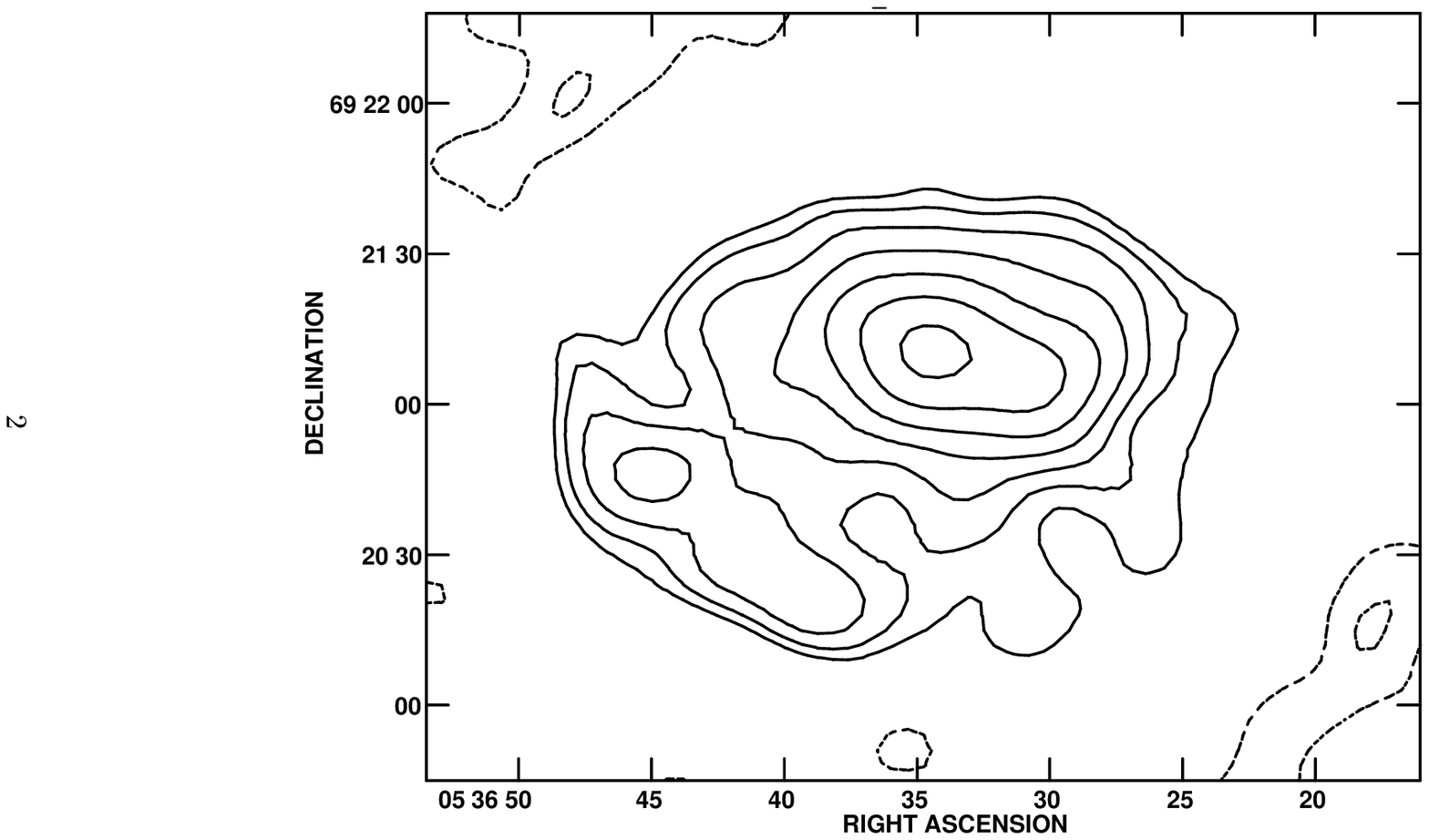,width=8.7cm,clip=,angle=270}
}
\vskip0.4cm
\centerline{\hoffset-0.5cm8.41~GHz\hskip7.5cm 15.4~GHz}
\hbox{
 \psfig{file=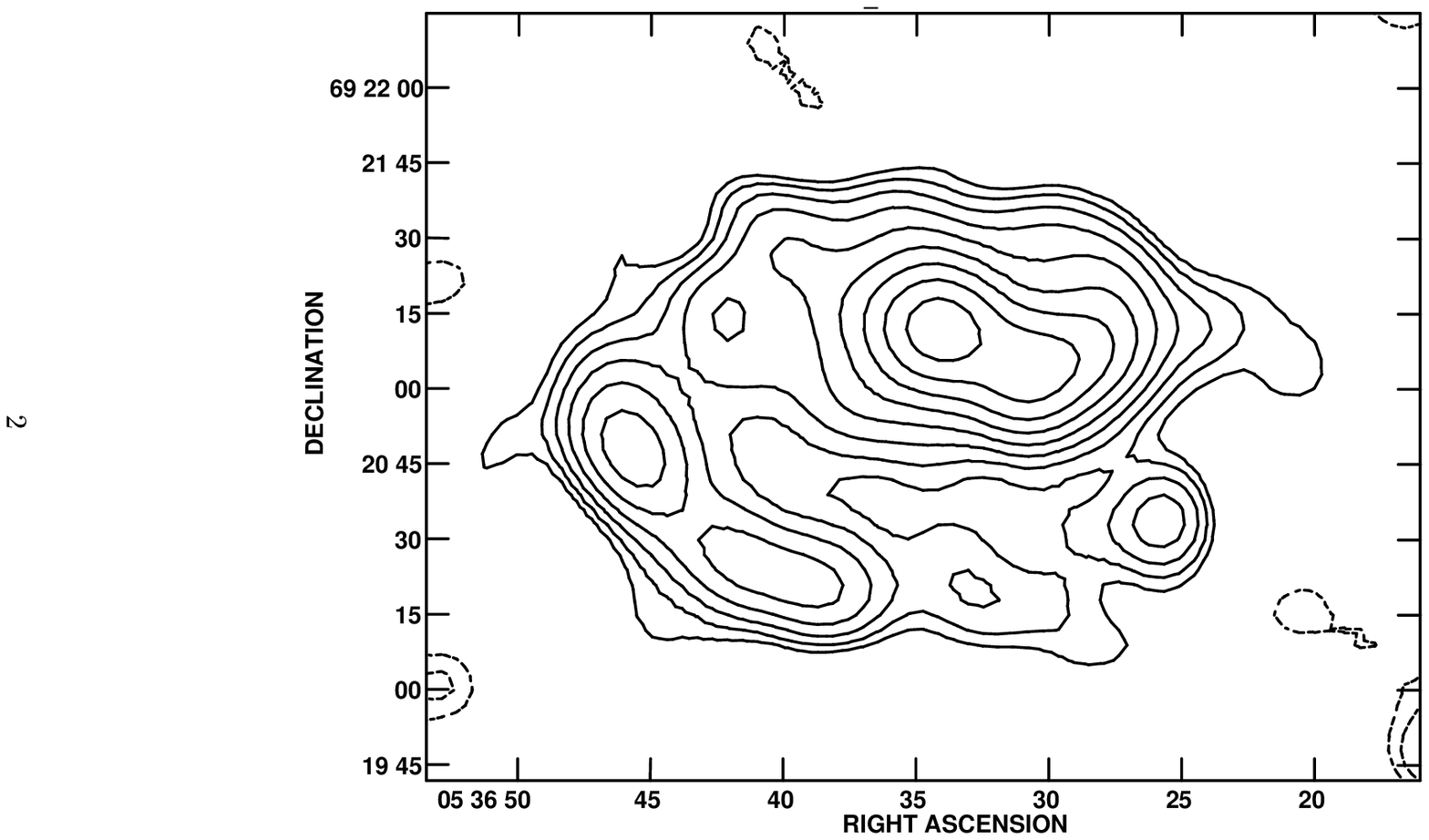,width=8.7cm,clip=,angle=270}\quad
 \psfig{file=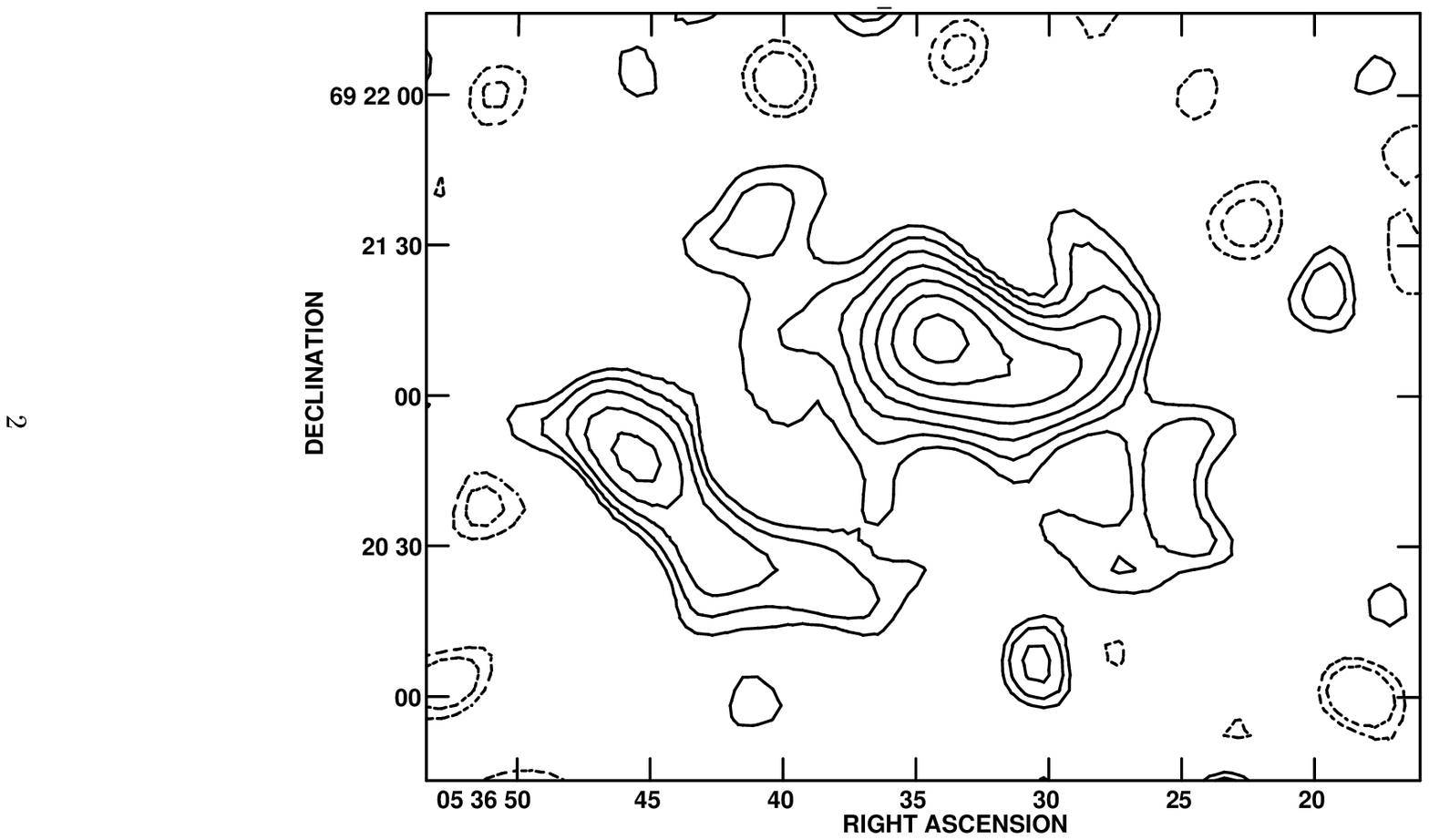,width=8.7cm,clip=,angle=270}
}
\caption[Maps of NGC~2146]{Maps of NGC~1961 at 1.46, 4.92, 8.41 and 15.4 GHz
at a resolution of 16  $\times$ 16 arcsec$^2$. 
The contour levels start at (-0.85, -0.60, 0.60, 0.85.....)  mJy beam$^{-1}$ 
(1.46~GHz and 4.92~GHz); (-0.28, -0.20, 0.20, 0.28,...)
mJy beam$^{-1}$ (8.41~GHz) and
(-0.48, -0.34, 0.34, 0.48....) mJy beam$^{-1}$ (15.4 GHz) and increase
by factors of $\sqrt{2}$. 
The rms noise levels are:
0.30 mJy beam$^{-1}$ (1.46 and 4.92~GHz), 
0.10 mJy beam$^{-1}$ (8.41 GHz) and
0.17 mJy beam$^{-1}$ (15.4 GHz).}
\end{minipage}
\end{figure*}

NGC~1961 was mapped with the VLA\footnote{
The VLA is operated by the National Radio Astronomy Observatory for 
Associated Universities Inc., under a cooperative agreement with the
National Science Foundation.
}
at 8.41~GHz and with the Cambridge 5-km
Ryle Telescope (RT) at 4.92 and 15.4~GHz.  Additionally, existing VLA data 
(B-array) at 1.46~GHz (Condon 1983) were available.  
When the observations of NGC~1961 were made (between 1990 and 1993), 
the RT had a band-width of
280~MHz split into 28 10-MHz frequency channels; together with the
minimum baseline of 18~m this provides excellent temperature sensitivity 
at both 4.92 and 15.4~GHz.

\begin{figure*}
\begin{minipage}{178mm}
\centerline{\hskip-0.5cm 1.46~GHz \hskip7.5cm    8.41~GHz} 
\hbox{
 \psfig{file=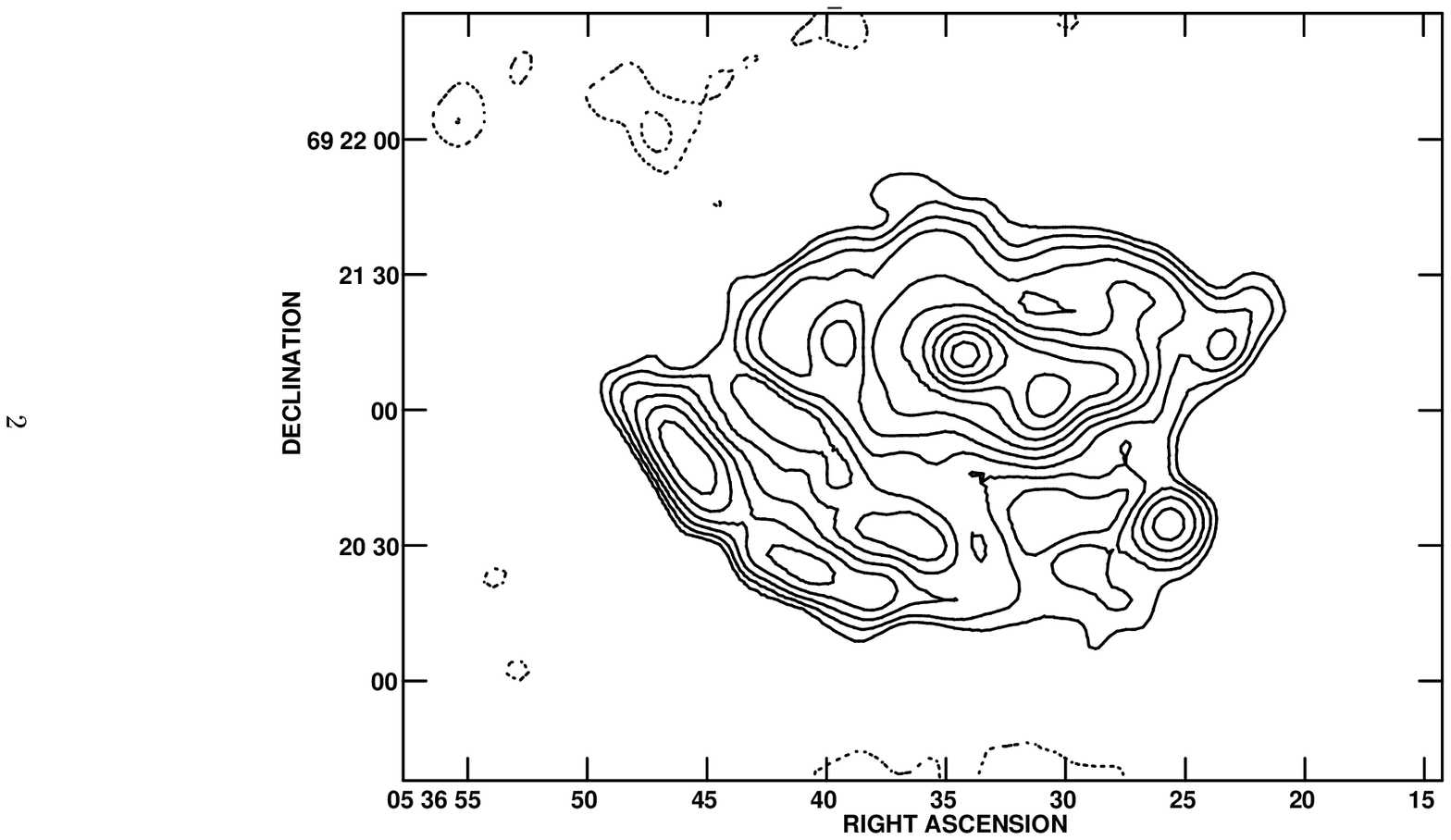,width=8.7cm,clip=,angle=270}\quad
 \psfig{file=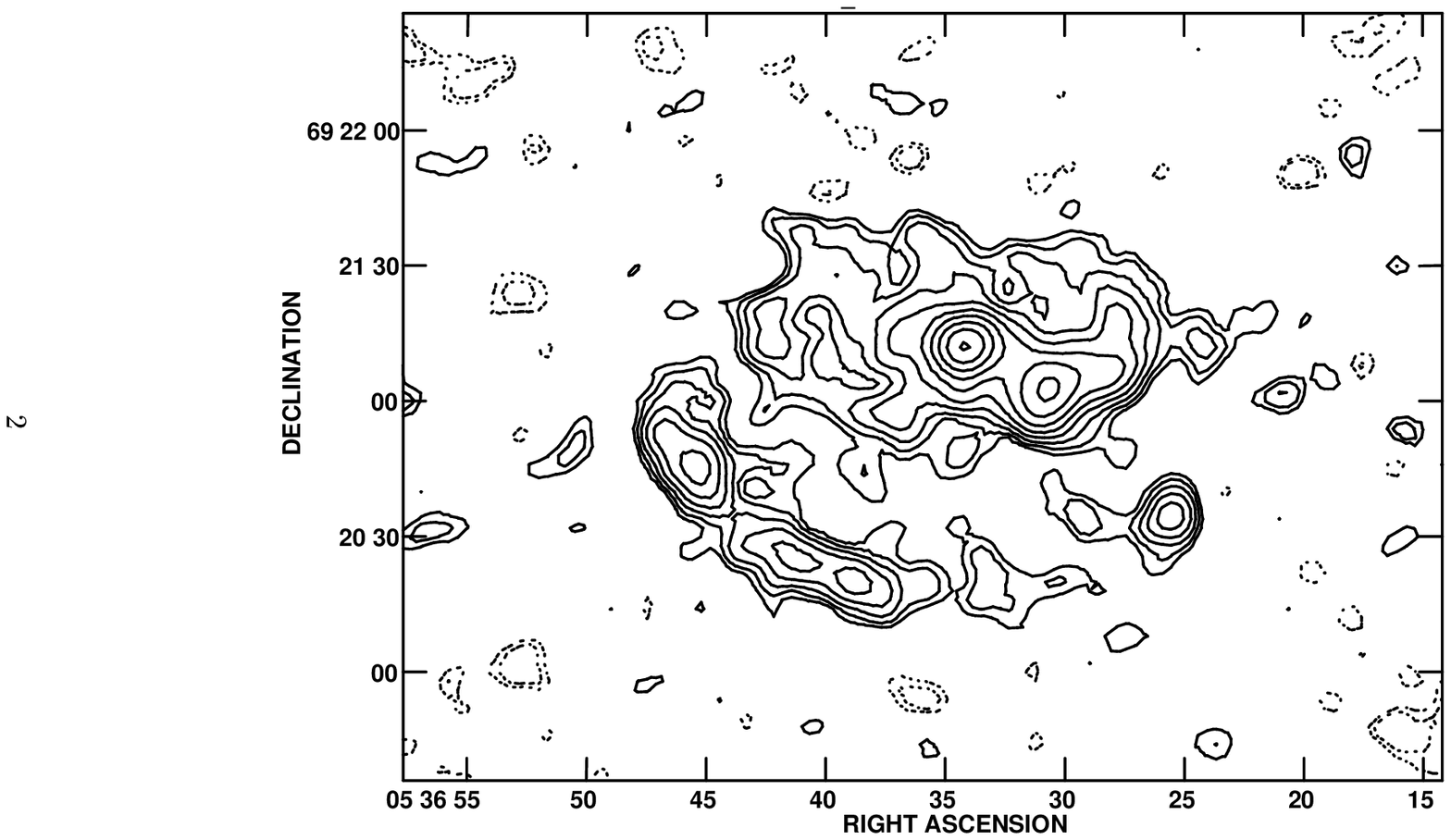,width=8.7cm,clip=,angle=270}
}
\caption[Maps of NGC~2146]{Maps of NGC~1961 at 1.46 and 8.41  at
a resolution of
8  $\times$ 8 arcsec$^2$. The contour levels start at (-0.48, -0.34,
0.34, 0.48 ....) mJy beam$^{-1}$ 
(1.46~GHz) and  at (-0.14, -0.10, 0.10, 0.14.....) mJy beam$^{-1}$ (8.41~GHz)
and increase by factors of $\sqrt{2}$. 
The rms noise levels are:
0.17 mJy beam$^{-1}$ (1.46~GHz), 
0.05 mJy beam$^{-1}$ (8.41 GHz).}
\end{minipage}
\end{figure*}

For the RT observations a total of five telescope configurations at each
observing frequency were employed giving a nearly fully filled aperture
out to a baseline of approximately 1 km.  A minimum of two 12-hour runs in each
configuration were obtained. This gave a resolution  of approximately 
$1.0 \times 1.0\mbox{\rm cosec}( \delta )$~arcsec$^{2}$ at 15.4 GHz and
$3.0 \times 3.0\mbox{\rm cosec}( \delta )$~arcsec$^{2}$ at 4.92 GHz.
At 4.92~GHz observations of phase-calibrators were made at the beginning
and end of each run, while at 15.4~GHz calibration observations were
interleaved with those of NGC~1961.  3C~286 and 3C~48 were observed
regularly as flux calibrators.
Calibration and data-editing was performed in the MRAO package 
{\sc postmortem}, with subsequent reduction
in {\sc aips} and the MRAO package {\sc anmap}.

The 8.41-GHz VLA observations were in the B, C and D arrays and were
carried out during 1990/91.  Reduction of the data followed standard VLA
procedures with calibration, editing and imaging performed in {\sc aips}.
The 1.46~GHz data were from Condon (1983) and no additional reduction
was required.
A number of maps were made at each frequency by tapering the aperture-plane
to highlight structure on different angular scales.

The data in this paper which are most seriously affected by the lack of short
baselines are those at 1.46 GHz, which are only sensitive to  structures on
scales smaller than about 2 arcmin. Very faint emission from the whole of 
the optical disc of NGC~1961 will not be adequately represented in our 
maps. The optical diameter of NGC~1961 is 4.6 arcmin; however the extent 
of the continuum emission at 1.46 GHz as determined by Shostak et al. (1982)
using the WSRT with a shortest baseline of only 36 m is approximately 
2.5 arcmin, with very little low-brightness emission outside that region.
A further test of the completeness of our mapping is to compare our 
measured flux densities with single dish data; these comparisons are 
reported in Table 1. It is apparent that to within the quoted uncertainties
our flux densities are consistent with those determined by other workers 
(including single dish measurements at 1.40~GHz and 4.85~GHz), 
although the best estimates represent 
a shortfall of 18\% of the flux density at 1.46~GHz and 12\%
at 4.92~ GHz. These results suggest that we are able to image the
extent of the continuum emission satisfactorily at all our observing 
frequencies.

So as to make the spectral index comparison as accurate as possible it is
crucial to ensure that the spectral index is determined from images all
of which are sensitive as far as possible to the same range of spatial 
structure. To achieve such a match, the UV-range of the data in the 
aperture plane used in the construction of the images above 1.46~GHz was 
restricted during the mapping programs so that the minimum and maximum
baselines (measured in wavelengths) used in the imaging were the same; this
involved removing baselines from all of the high-frequency data. While it is
not possible to ensure that the precise sampling of the aperture plane 
is the same in all cases, for both the VLA and RT the dense sampling near
the shortest baselines leads to similar overall coverage of the UV plane.
Furthermore the use of CLEAN does, to some extent, compensate for the 
resulting marginally different beamshapes. A possible problem results from 
the fact that the extended, low-brightness structure often has a steep
radio-frequency spectrum. Missing the short baselines could lead to 
a negative region (``negative bowl'') around the source and hence
a (frequency-dependent) error in the zero level of the image. The final 
images were checked for the existence of such an effect and in 
each case any offset of the local zero level was found to be less than the
thermal noise (this is clear from Figs. 1 and 2 where no evidence of a 
significant negative bowl exists). We are confident that the procedures
that we have adopted here are sufficient to reduce any artefacts resulting
from the sampling of the aperture to levels significantly below those
of the thermal noise. 

%

All the maps were {\sc CLEAN}ed and a set of maps at a common resolution of
$16 \times 16$~arcsec$^{2}$ were used for the multi-frequency
comparison (at 1.46~GHz this was achieved by convolving the supplied image).
All maps have been corrected for the primary beam response of the
telescope used to make the observations.

\begin{table}
\caption{Integrated flux densities}
\begin{tabular}{ccc}
\hline
Frequency &  Flux Density & Reference \\ 
   $\rm [MHz]$  &    $\rm [Jy]$    &        \\
\hline
38    & 3.40 $\pm$ 0.85 & Howarth (1990) \\
57.5  & 2.0 $\pm$ 0.5  &  Israel \& Mahoney (1990) \\
151   & 1.17 $\pm$ 0.21 & Howarth (1990) \\
1400  & 0.191  & White \& Becker (1992) \\
1465  & 0.149  $\pm$0.015  & this work \\
1490  & 0.182 $\pm$ 0.023  &  Condon (1987) \\
4850  &  0.057 $\pm$ 0.009 & Becker, White \& Edwards (1991)\\
4919  &  0.050 $\pm$ 0.006  & this work \\
8414   & 0.028 $\pm$ 0.003 & this work  \\
10550  & 0.031 $\pm$ 0.005  & Niklas et al. (1995)\\
15360  & 0.019 $\pm$ 0.002 & this work \\ 
\end{tabular}
\end{table}

\section{Results}

\subsection{Radio images}

Fig. 1 shows the radio maps at the four frequencies at
a resolution of 16 arcsec; 
maps at a resolution of 8~arcsec 
are shown in Fig. 2. 
In Fig. 3 the 1.46~GHz radio map is overlaid on
an optical  image obtained from the Digitized Sky Survey produced
by the Space Telescope Science Institute.
The radio images at all frequencies
show a peculiar morphology which are similar to the optical image. 
The bright central nucleus is resolved in the higher resolution images and
consists of two peaks -- one which corresponds to the optical peak and 
a second peak which coincides with a dust lane.
Furthermore, a bright radio arm is visible in  the south-east, coinciding
with the  bright optical south-eastern arm. The optical spiral arm
extending from of the west of the nucleus and to the south is also 
visible in the radio image. At all frequencies, except for 15.4 GHz,
there is an extended envelope of radio emission visible which
surrounds the nucleus and the south-eastern arm and which contains, 
at 1.46 GHz,  about 50$\%$ of the total radio flux.

\begin{figure}
 \psfig{file=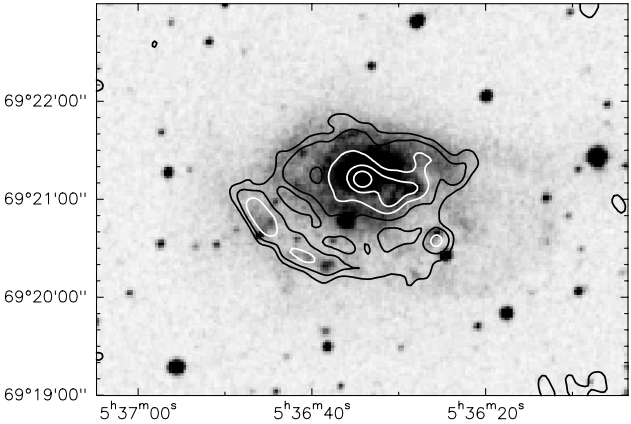,width=9.2cm,clip=,angle=270}
\caption[Maps of NGC~2146]{The radio map at 1.46 GHz from Fig. 2
overlaid on an optical image. The contour lines start at 0.34 mJy and increase
by factors of 2.}
\end{figure}

\subsection{Radio spectrum and spectral index distribution}

Integrated flux densities at the four observing frequencies were determined
from the low-resolution maps and in Table 1 the results are summarized 
together with flux densities taken from the literature. 
The integrated spectrum is shown in Fig. 4; the spectrum is well-fitted by a 
power law with a spectral index $\alpha = 0.85$ 
($S(\nu) \propto \nu^{-\alpha}$), which is not unusual for a spiral galaxy;
the average spectral index for a sample of spiral galaxies 
is 0.74 $\pm 0.03 $ (Gioia et al. 1982). 

As discussed in Section 2 there may be some flux missing from our images
due to the restricted UV-coverage on the shortest baselines; our estimate
of this is 18\% (33 mJy) at 1.46 GHz and 12\% (7 mJy) at 4.92 GHz. The spectral
comparison has been performed using images with matched UV-coverage and 
resolution so that the spectra of structure of similar spatial scales may be 
compared. If the emission is regarded as features superimposed on a 
background of emission, then on those scales which are properly represented 
in our imaging the spectral results will be accurate. A problem
may exist, however, in those regions which do not show significant
structural variation. In this case we can quantify the likely errors. If
we take a worst case of 33 mJy missing flux density at 1.46 GHz distributed
over an area of 9 armin$^2$ (slightly larger than the observed mappable extent
of the source), then this would correspond to an error of 0.26 mJy/beam
over the source at 1.46 GHz, which is less than the 
thermal noise and at a level of
approximately 40\% of the lowest contour in Fig. 1. Since we would expect
the spectrum of the most extended material to be steep,
this effect becomes less important at the higher frequencies. 
The effect of the missing short spacings is therefore in the worst case to
underestimate very slightly the flux  density at 1.46 GHz relative to
the higher frequencies and therefore to underestimate also the spectral 
index. However, this effect will lead to a systematic error which is 
considerably smaller than the statistical error due to the thermal noise
and in what follows it will be ignored.


\begin{figure}
\psfig{file=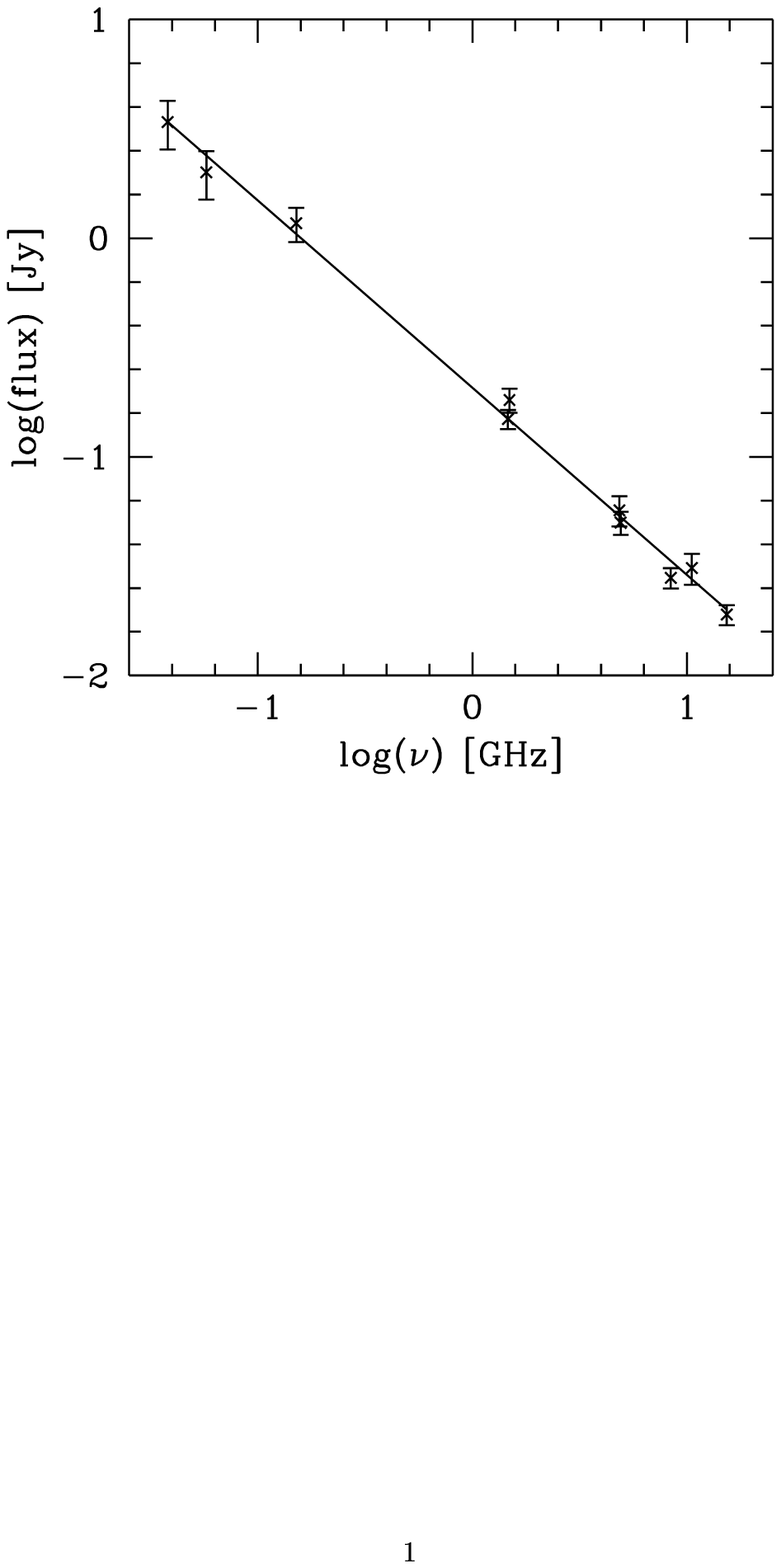,width=8.cm,clip=,angle=0}
\caption[Spectral index map]{
The radio spectrum of NGC~1961. The data are listed in Table 1.
The line is the best-fit power-law spectrum and has a slope of 0.85.
}
\end{figure}

\begin{figure*}
\begin{minipage}{178mm}
\centerline{\hskip 4.cm $\alpha_{1.46-4.92{\rm GHz}}$ \hskip7.cm
 $\alpha_{1.46-8.41{\rm GHz}}$\hfill}
\hbox{
 \psfig{file=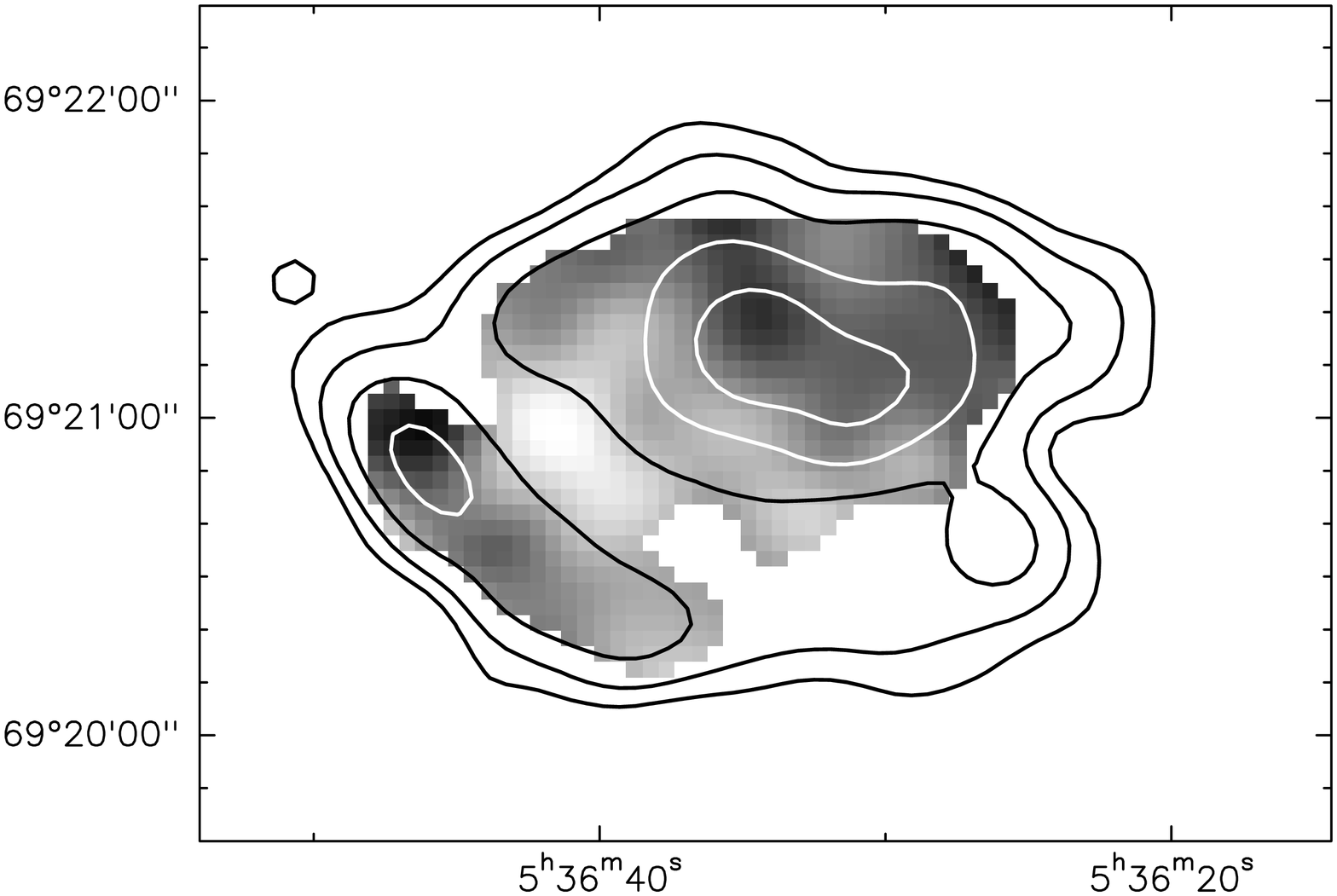,width=8.7cm,clip=,angle=270}\quad
 \psfig{file=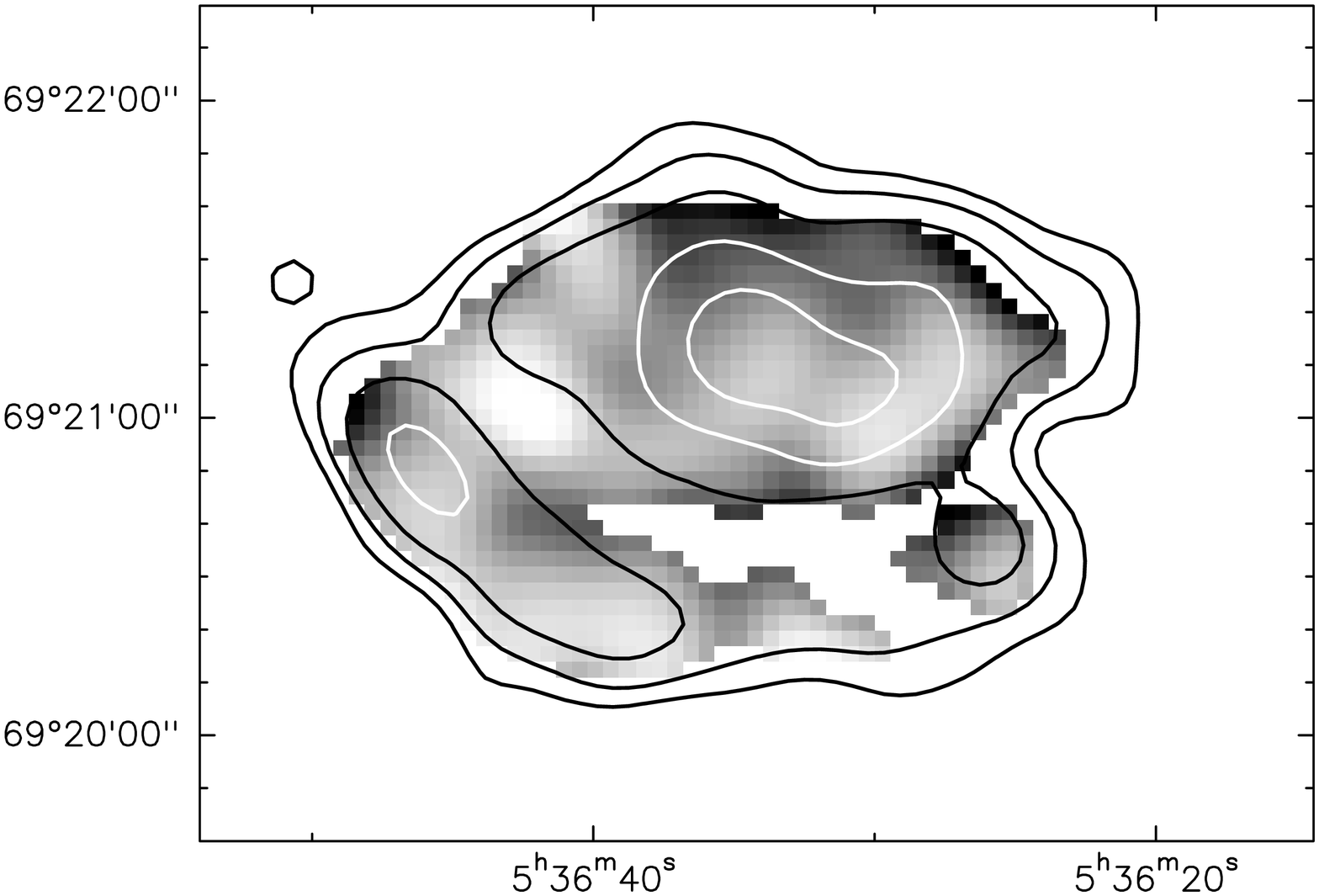,width=8.7cm,clip=,angle=270}
}
\caption[Maps of NGC~2146]{Maps of the spectral index
between 1.46 and 4.92~GHz (left) and between 
1.46 and 8.41~GHz (right) at a resolution of 
16  $\times$ 16 arcsec$^2$. The greyscales range from
0.1 to 1.2 ($\alpha_{1.46-4.92GHz}$) and from 0.6 to 1.2 
($\alpha_{1.46-8.41GHz}$). Overlaid is a 1.46~GHz image with
contour lines starting at 0.6 mJy and increasing by factors of 2.}
\end{minipage}
\end{figure*}

Fig. 5 shows spectral index maps between 1.46 and 4.92 GHz
and between 1.46 and 8.41 GHz, each at 
16-arcsec resolution and gated  so that the spectral index is only 
calculated where the flux density exceeds 3$\sigma$ at both frequencies.
The spectral index between 1.46 and 4.92 GHz ($\alpha_{1.46-4.92\rm GHz}$)
shows a peculiar behaviour in the sense that at the maxima of the radio 
emission
(i.e. at the nucleus and  on the south-eastern arm)
the spectral index is steep.  
Away from these sites the spectral index flattens, except
towards the north and west of the nucleus and towards the north of the arm.
The behaviour of $\alpha_{1.46-8.41\rm GHz}$ is less uniform which is
largely due to the contribution of thermal emission at the higher frequency. 
The most noticeable flattening of  the spectral index  in Fig. 5 is 
between the nucleus and the south-eastern arm where
$\alpha_{1.46-4.92\rm GHz}=0.2\pm 0.3$  and 
$\alpha_{1.46-8.41\rm GHz}=0.55\pm 0.2$.
The spectral index variation that would normally be expected due to
synchrotron and inverse Compton aging of the CR electrons 
is a gradual steepening of the spectrum away from the
sites of acceleration. However,
at the distance of NGC~1961 our 16-arcsec beam corresponds to 6.4 kpc which
is larger than the typical scale of electron diffusion, and
we are effectively averaging over a CR electron population of varying ages.

\section{Separating the thermal and synchrotron emission}

The maps at 4 frequencies allow us to separate the radio-continuum emission
into thermal and nonthermal components.
A spectral-fitting technique is used to estimate the two contributions and
determine spatial variations in the form of the synchrotron spectrum.
For a set of maps at a common resolution a function of the following
form is fitted at each pixel
\begin{equation}
I(\nu)=SC(x_{\rm B},\gamma)+T(\nu_0)\nu^{-0.1},
\end{equation}
where $T(\nu_0)\nu^{-0.1}$ is the emission from
optically thin thermal Bremsstrahlung and $SC(x_{\rm B},\gamma)$ is the 
synchrotron emission from an aged electron population with injection index
$\gamma$ which has been radiating for a time $t$;
$x_{\rm B}=\nu/\nu_{\rm B}$, where $\nu_{\rm B}$ is the break frequency.
During the fitting procedure the minimum in $\chi^2$ is found by 
varying $S$, $\nu_{\rm B}$ 
and $T(\nu_0)$ with the additional constraint that they are positive definite.  
The form of the synchrotron spectrum $C(x_{\rm B},\gamma)$
depends on many processes, but to
simplify the problem we make the following approximations.
\begin{enumerate} 
\item The dominant electron energy loss mechanisms are synchrotron  
and inverse Compton emission. These
energy losses are proportional to the energy density of the magnetic
field, $\ub$, and the energy density of the radiation field,
$\urad$, respectively. 
The magnetic field can be estimated from the standard minimum energy
assumption. With a high-frequency cut-off
of 10~GHz, a low-frequency cut-off of 100~MHz, a spectral index of
0.8, and a ratio of CR proton-to-electron energy density of $k=100$, this yields $B=15\mu$G.
$\urad$ can be estimated from the bolometric luminosity, $\lbol$, 
of the galaxy by assuming that the
radiation emitting matter is distributed  in a flat cylinder:

\begin{equation}
\urad={2 L_{\rm bol} \over c\pi R^2}=
{\bigl(L_{\rm bol}/ {\rm W}\bigr) \over \bigl(R/{\rm kpc}\bigr)^2}
\times 1.4 \times
   10^{-35}\biggl[{\rm eV \over cm^3}\biggr].
\end{equation}

We approximate  $\lbol$ as the sum of the far-infrared and blue emission 
(yielding $\lbol=1.3\,10^{38}$ W) and hence find $\urad=1.5$ eV cm$^{-3}$. 

\item Diffusion in the disc of the galaxy is negligible on scales
larger than our observing beam, which in this
case is a good approximation.

\item The electrons which are radiating within a given resolution element
were all accelerated (injected) a time $t$ ago (the electron age).

\item The electron loss time is much greater than the pitch angle randomization
time --- the so-called JP model (Jaffe \& Perola 1974).

\item The electron injection index is $\gamma= 2.0$.
\end{enumerate}

With these assumptions the electron energy distribution with an
injection index $\gamma$ is given by:
\begin{equation}
N(E,t) = \left\{ \begin{array}{ll}
N_{0}E^{-\gamma} (1 - \epsilon (\ub+\urad) E t)^{\gamma -2} & E < E_{B} \\
0 & E \ge E_{B} \\
\end{array} \right.
\end{equation}
with $\epsilon=(4/3)(\sigma_T c/mc^2)$, $\sigma_T$ being the Thompson
cross section, and $E_B=(\epsilon(\ub+\urad)t)^{-1}$.
The synchrotron spectrum is obtained by convolving this
electron energy distribution with the spectrum for a single electron.
The break frequency is given by:
\begin{equation}
\nu_{\rm B} =(3/2) \nu_G E_{\rm B}^2 \propto
\frac{B}{(\urad+\ub)^2t^2}
\end{equation}
where $\nu_G$ is the gyrofrequency.
\begin{figure*}
\begin{minipage}{178mm}
\hbox{
 \psfig{file=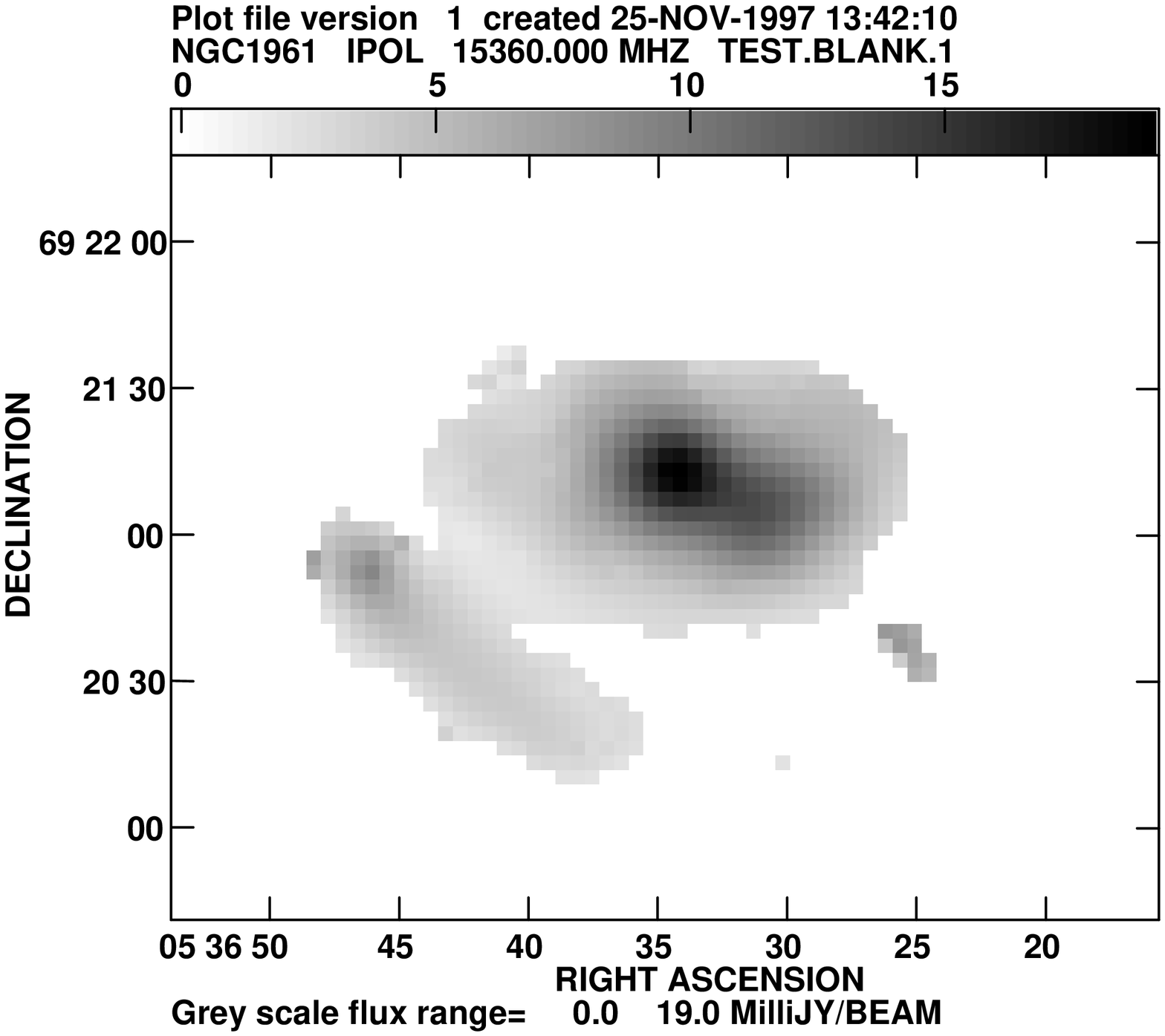,width=5.5cm,clip=,angle=270}\quad
 \psfig{file=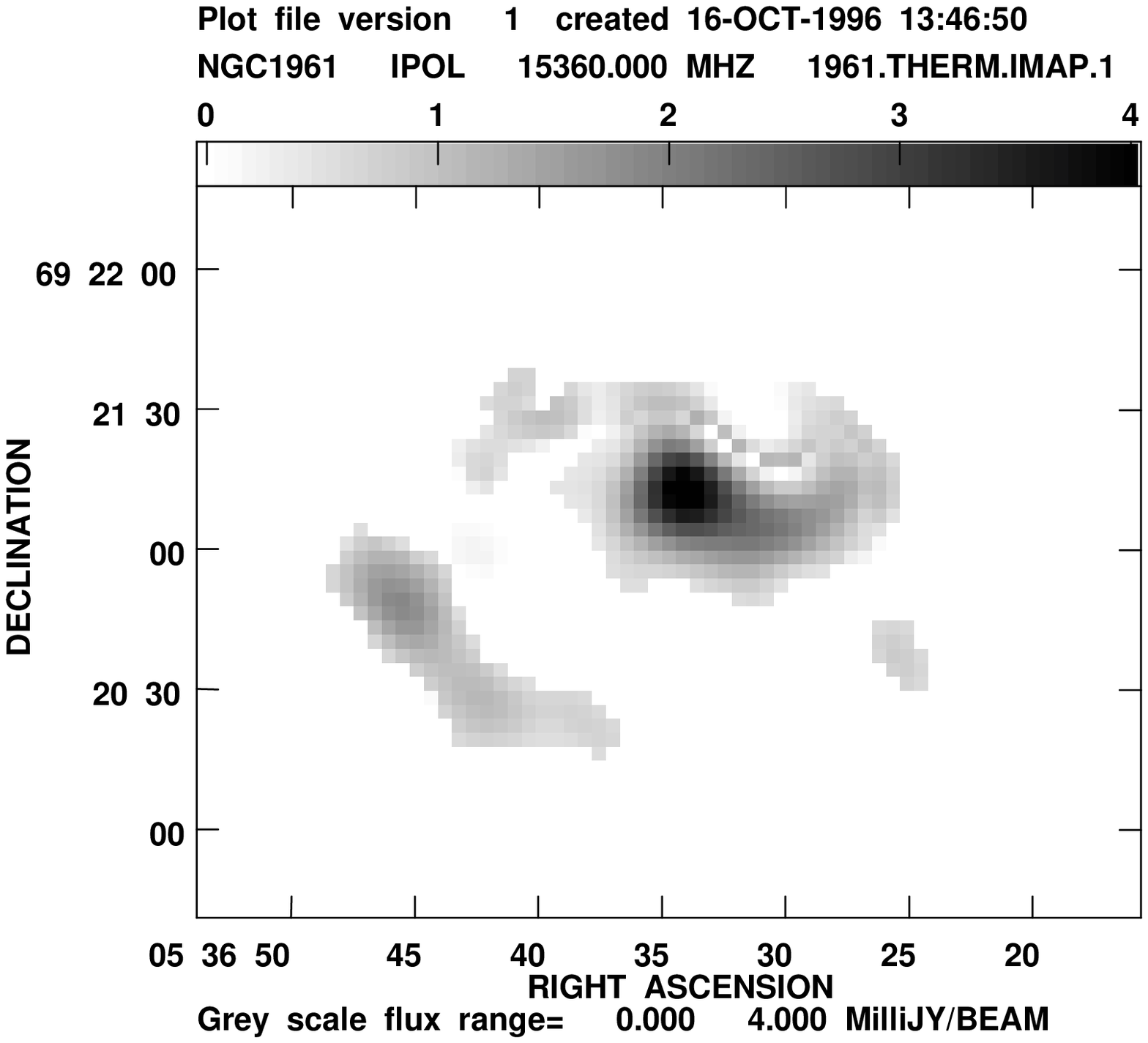,width=5.5cm,clip=,angle=270}\quad
 \psfig{file=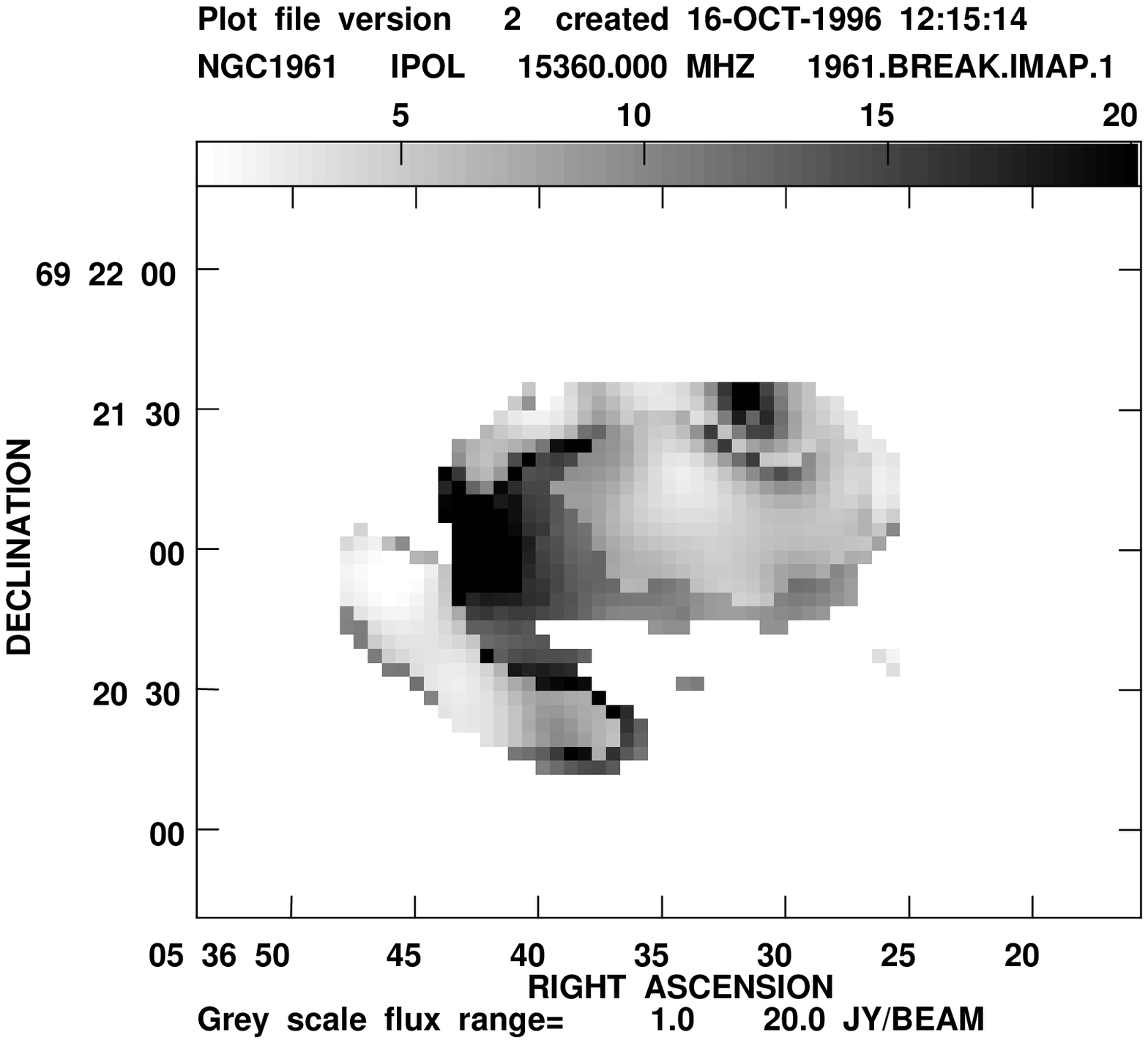,width=5.5cm,clip=,angle=270}\quad
}
\caption[Maps of NGC~2146]{Maps of the synchrotron emission (left),
thermal emission (middle) and break frequency (right) from which the
age of the electrons can be 
derived. For the synchrotron emission
the greyscale ranges from 0 (white) to 19 mJy/beam (black),
for the thermal emission from 0 (white) to 4 mJy/beam (black) and for
the break frequency from 1 (white) to 20 GHz (black) (corresponding to
 electron ages from $1.4\,10^7$ yr (white) to $3\,10^6$ yr (black)).
}
\end{minipage}
\end{figure*}
The fitting procedure was applied to all pixels at which the
flux density  exceeded 3 times the noise at 1.46, 4.92 and
8.41~GHz and 1.5 the noise at 15.4~GHz. 
The lower gate value at 15.4 GHz was chosen in order included
the extended, predominantly nonthermal, radio emission in the analysis.
Since the fitting procedure takes full account of the error in the data,
the quality of the separation is not affected by this. 
In Fig. 6 the results of this separation are shown in  the
form of maps of the thermal emission, synchrotron emission and
electron age derived from the breaks in the synchrotron emission.
The synchrotron emission is more extended than the thermal emission
which is mainly concentrated around the maximum of the radio emission 
and the south-eastern arm. The electron age reflects the peculiar 
behaviour which
is seen in the spectral index maps and noted above. The ages are
highest at the maxima of the radio emission where the 
spectrum between 1.46 and 4.92~GHz is steepest. 
The total thermal radio continuum emission found from this 
fitting procedure is 16 mJy at 1~GHz (i.e. $\sim 10 \%$ of the total emission).

\begin{figure*}
\begin{minipage}{178mm}
\centerline{\hskip3.cm 1 \hskip5cm  2 \hskip5cm  3 \hfill} 
\hbox{
 \psfig{file=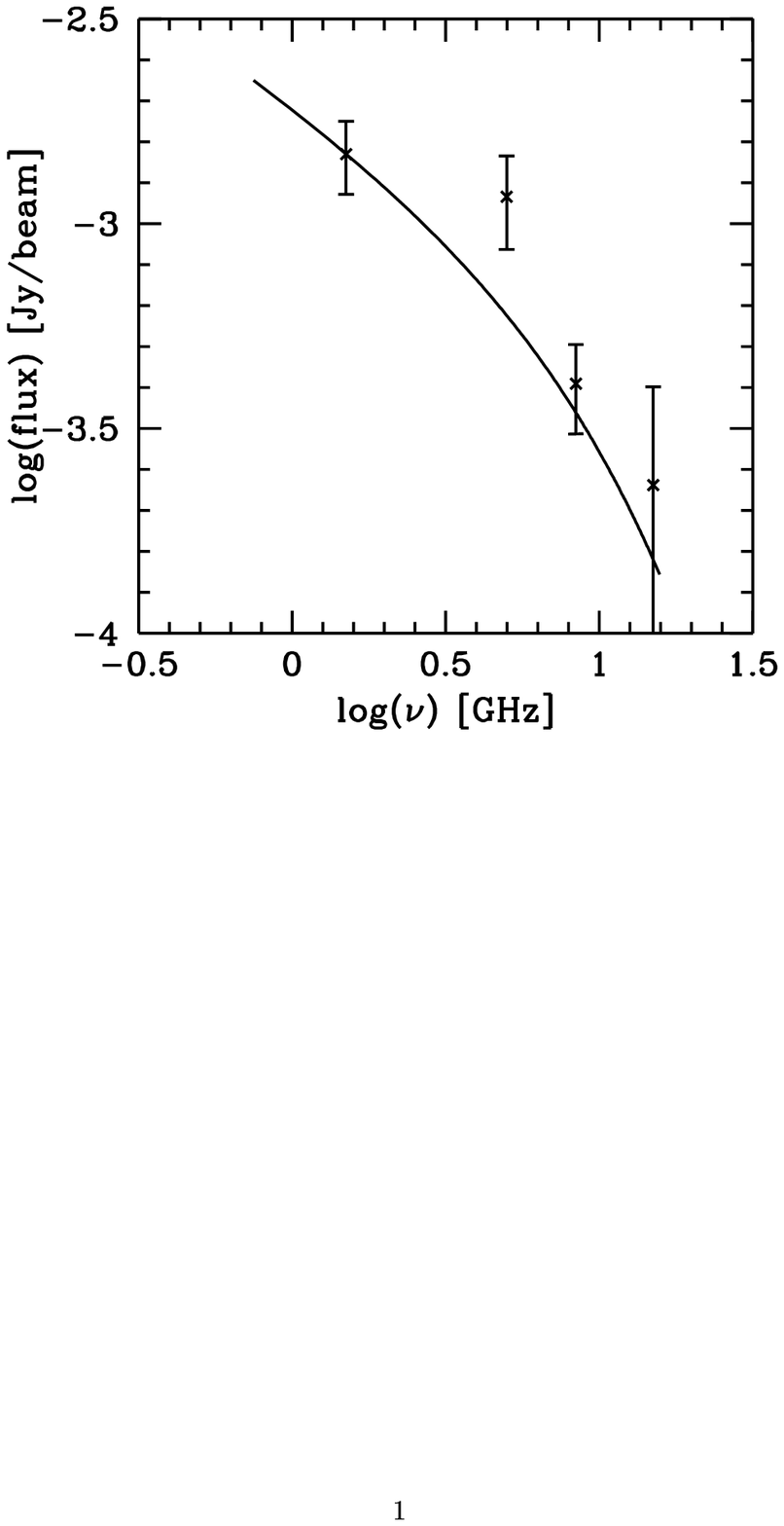,width=5.cm,clip=,angle=0}\quad
 \psfig{file=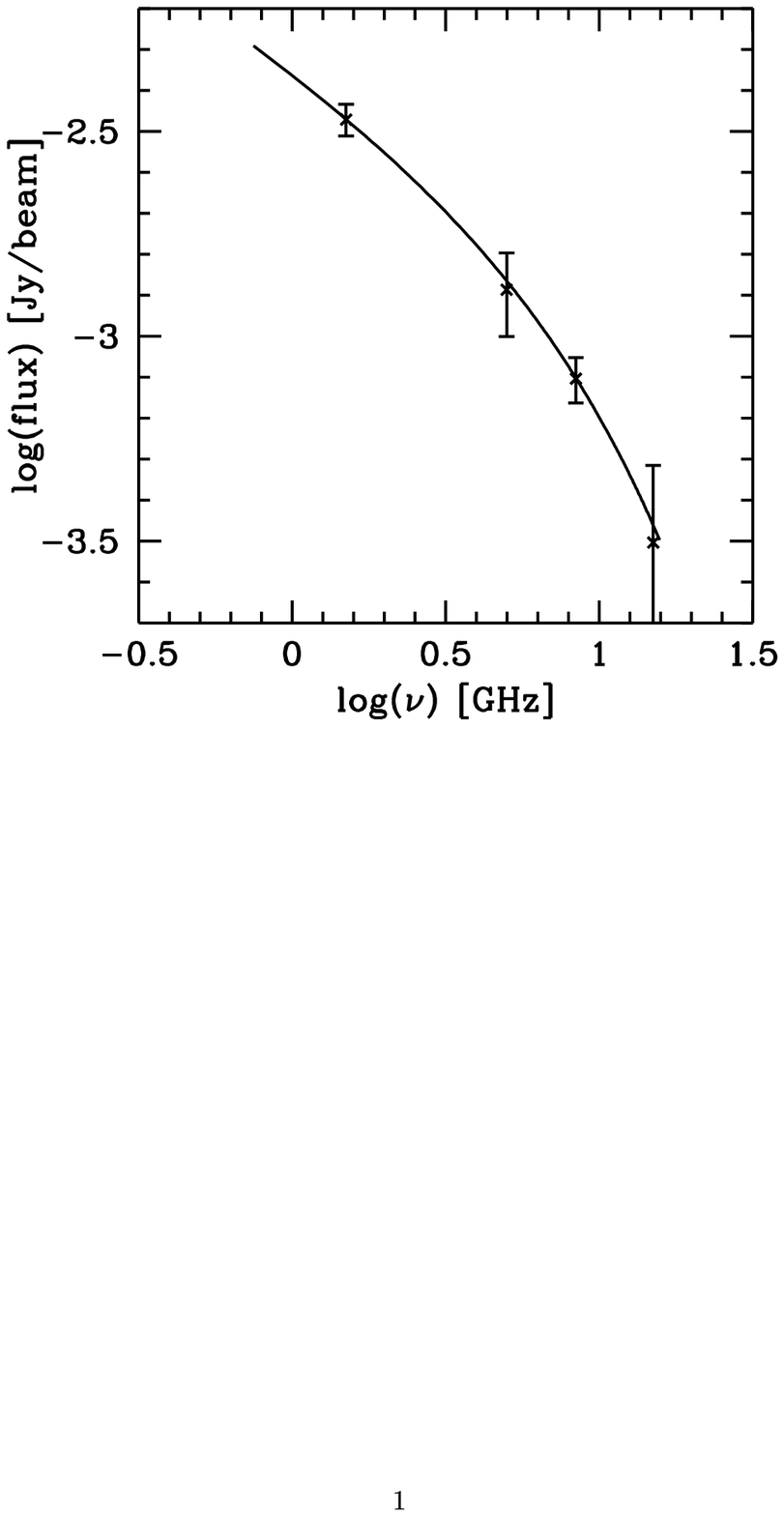,width=5.cm,clip=,angle=0}\quad
 \psfig{file=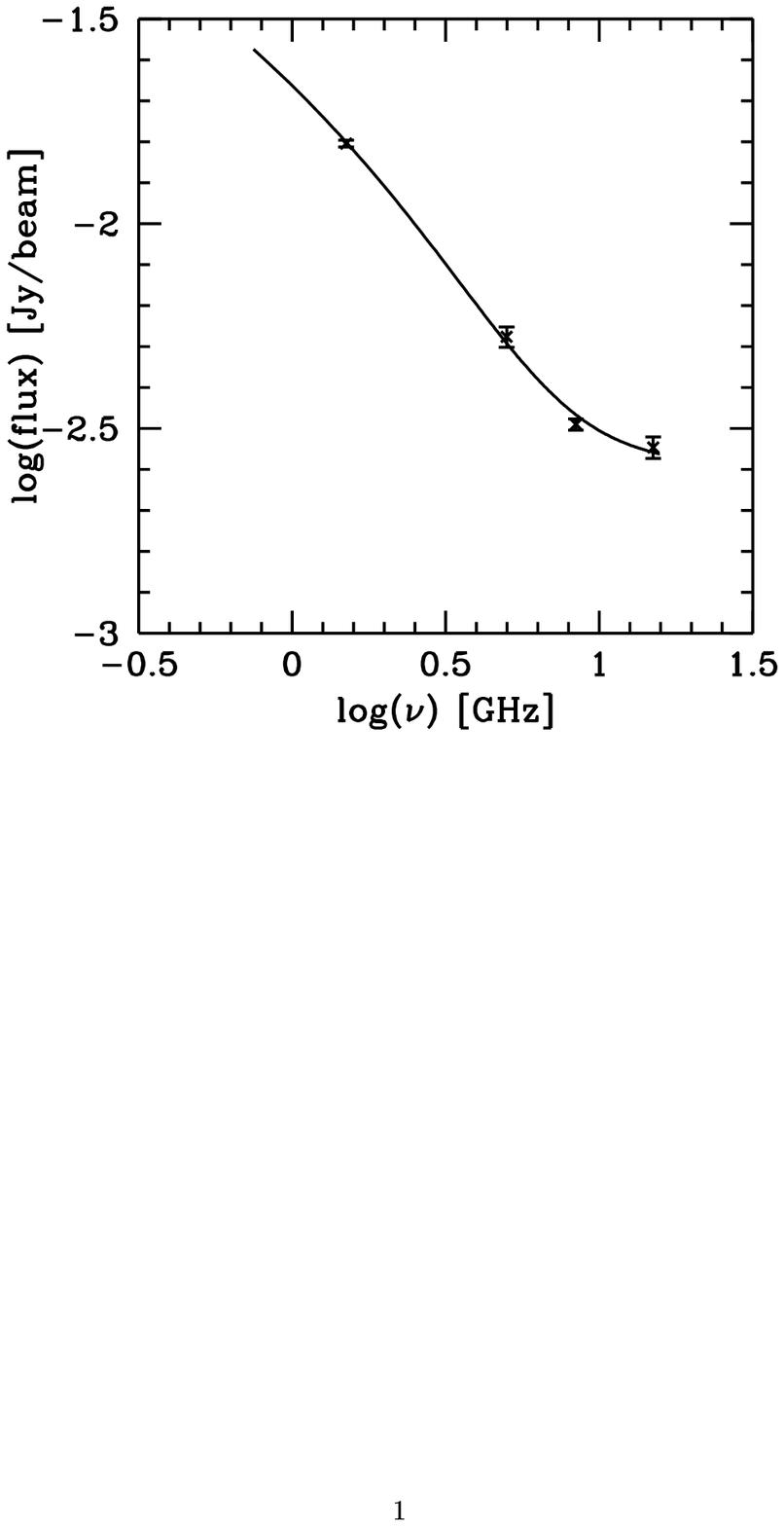,width=5.cm,clip=,angle=0}\quad
}
\vskip0.2cm
 \centerline{\psfig{file=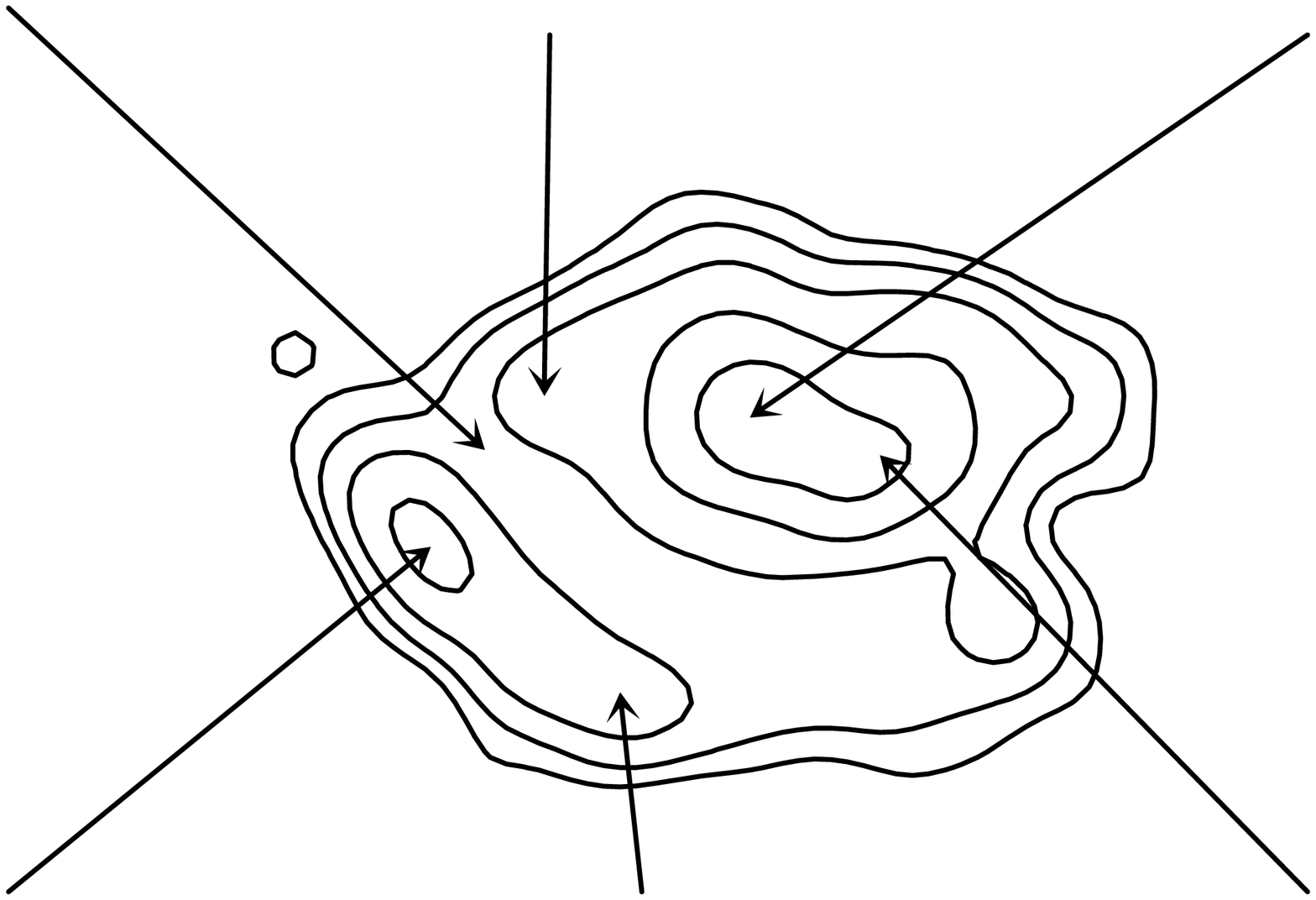,width=4.5cm,clip=,angle=270}}
\vskip0.2cm
\centerline{\hskip3.cm 4\hskip5.cm 5\hskip5cm   6 \hfill} 
\hbox{
 \psfig{file=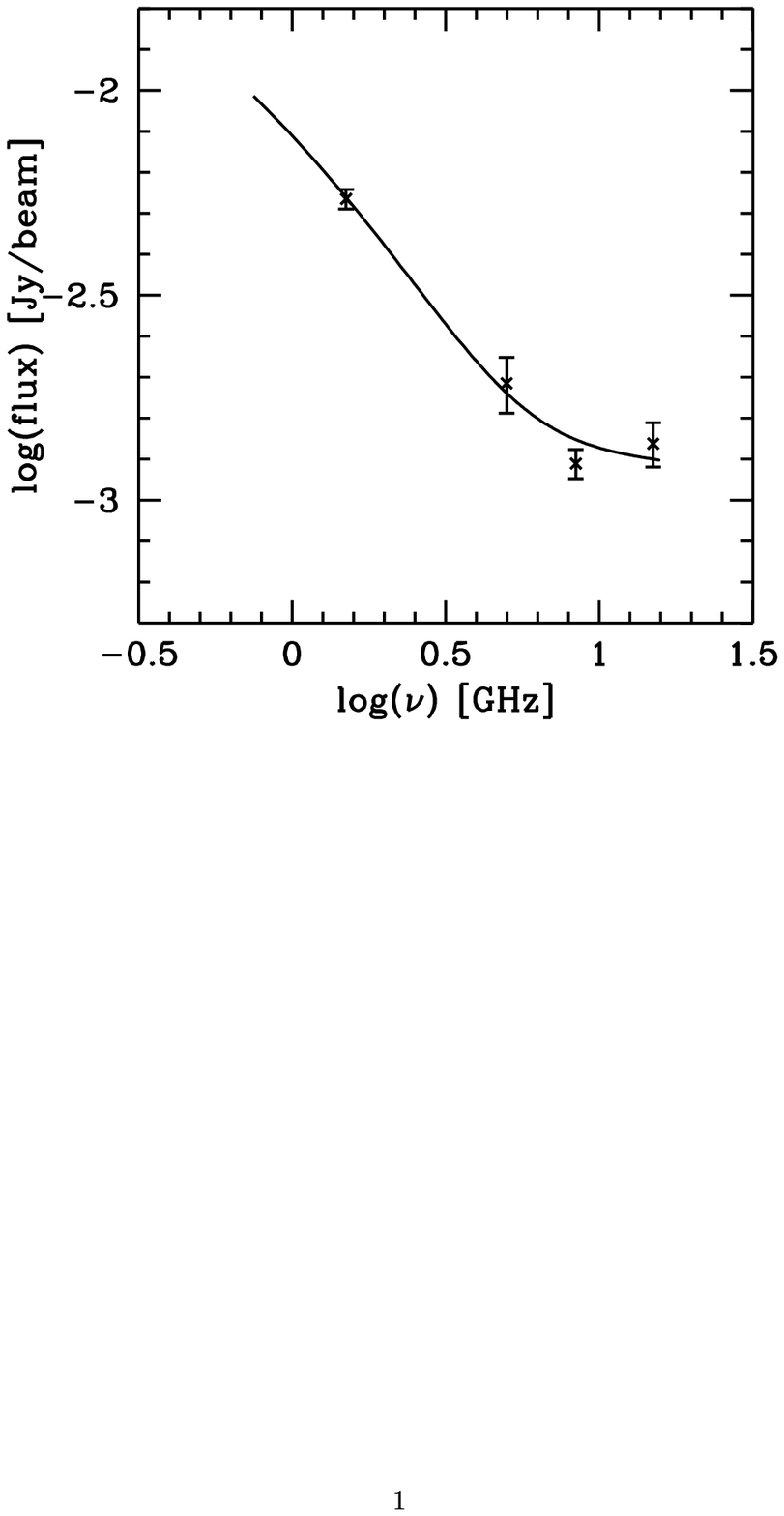,width=5.cm,clip=,angle=0}\quad
 \psfig{file=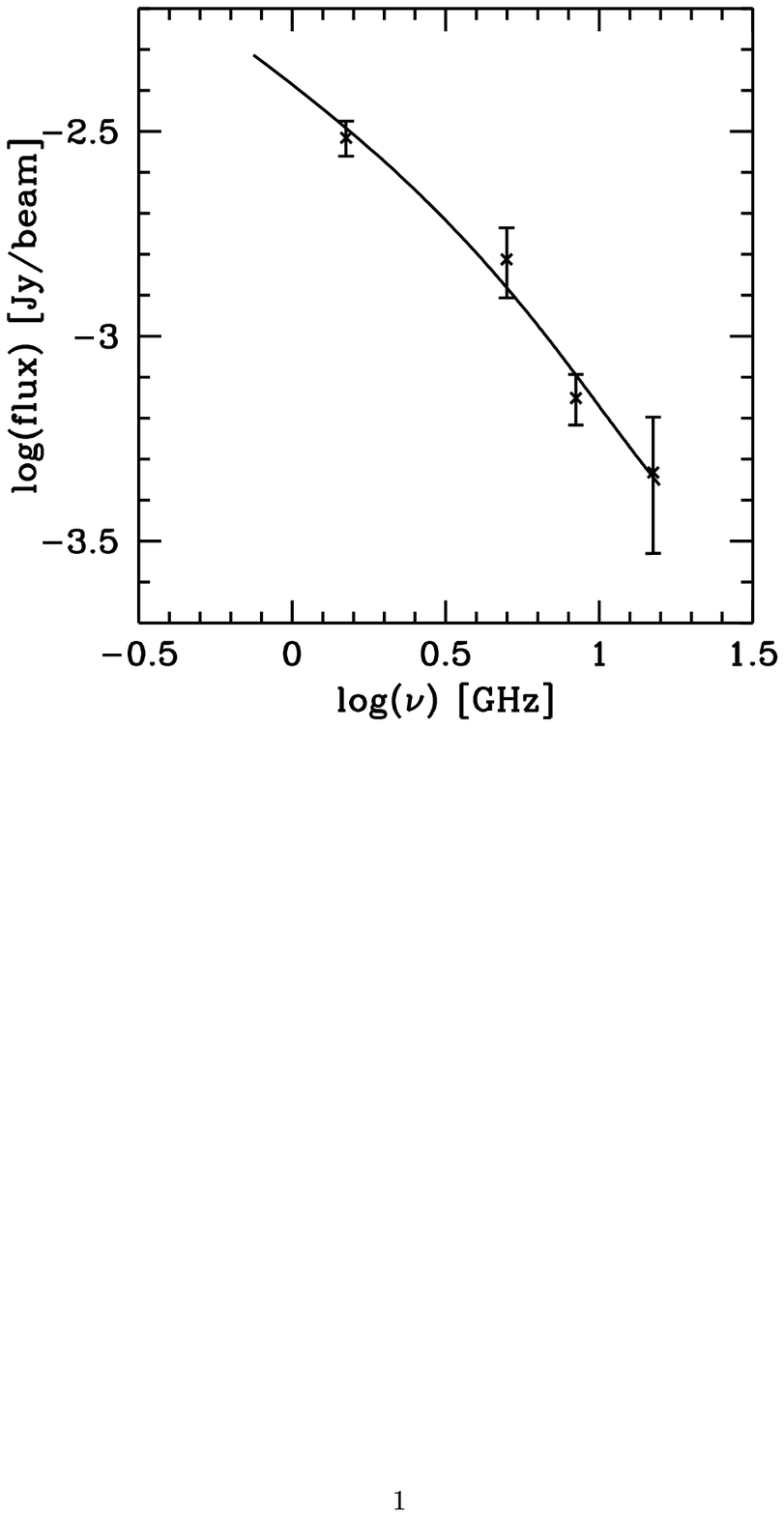,width=5.cm,clip=,angle=0}\quad
 \psfig{file=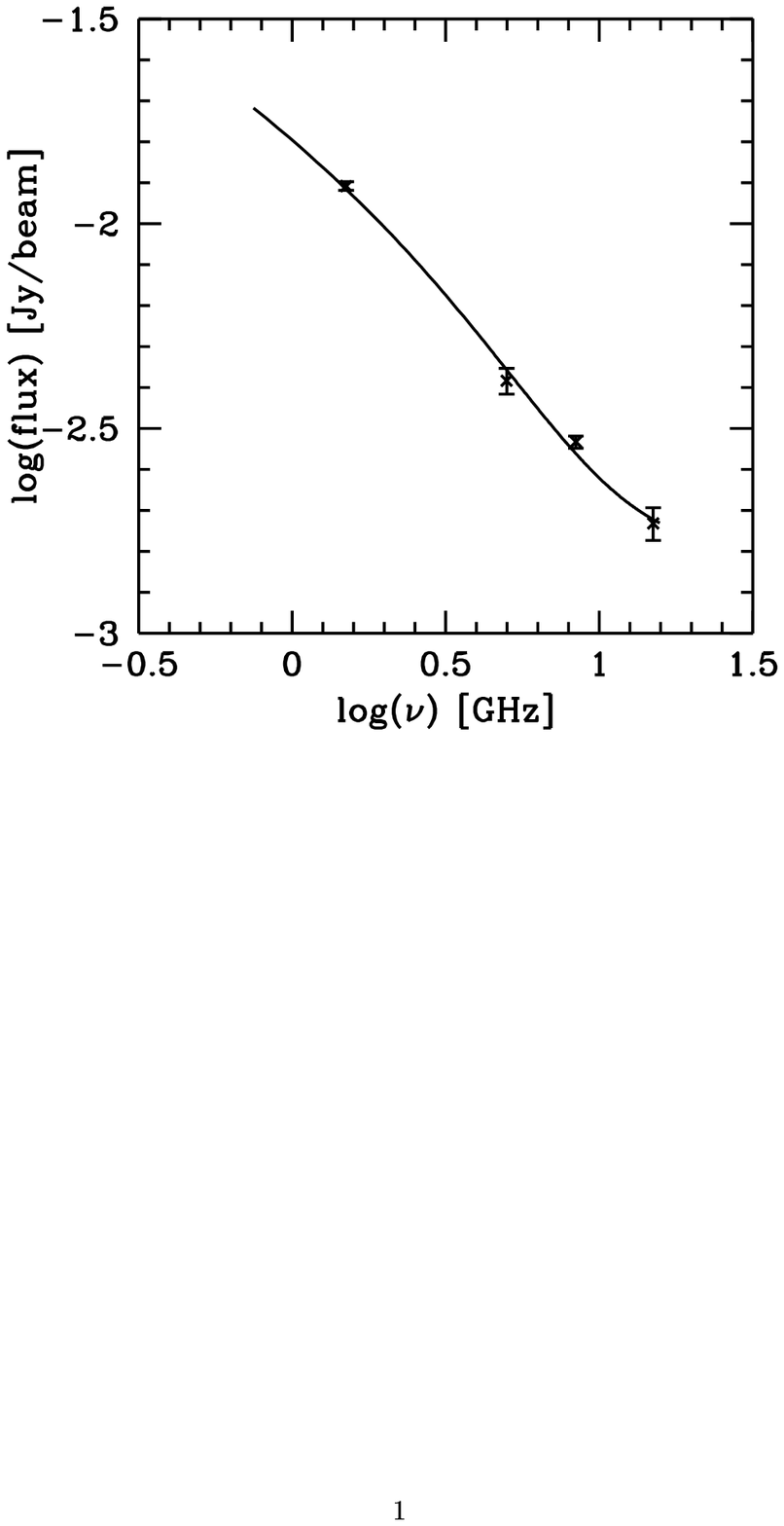,width=5.cm,clip=,angle=0}\quad
}
\caption[Maps of NGC~2146]{
Selected spectra in NGC~1961 together with 
the positions where they were extracted relative to a 1.46~GHz
image.}
\end{minipage}
\end{figure*}
Fig. 7 shows selected radio-continuum spectra of NGC~1961 and the fits 
associated to
these data together with their positions relative to a 1.46~GHz map.
The steep spectrum at the nucleus and at the maximum of the south-eastern
spiral arm can be seen (positions 3 and 4). At these locations the
flattening of the spectra towards higher frequencies due to
a high thermal component is clearly visible and the thermal emission
can be determined with good accuracy. The thermal fraction
is relatively high (25$\%$ at 1~GHz) and the nonthermal spectrum steep 
($\anth = 1.7$  between 1.46 and 4.92 at both positions). 
The derived break frequency
is low, suggesting that acceleration has not taken place for a considerable period.
Away from these positions, the radio spectra show less flattening
towards higher frequencies (positions 5 and 6), 
or none at all (position 1 and 2) and therefore
indicate a lower thermal contribution. 
The radio spectra at the positions 1 and 2 are  flat at low
frequencies and steepen toward higher frequencies. 
The break frequency
is high, indicating recent acceleration of CR electrons.

\section{Discussion}

\subsection{FIR and radio luminosity}

\begin{table}
\caption{Global properties}
\begin{tabular}{lcc}
Distance    & 82.9 Mpc     & SHSVV$^1$ \\
$D_{25}$ &  4.6 arcmin  & RC3\\
Inclination & $50^{\degr}$ & SHSVV \\
$P_{1.49GHz}$ & 1.5 10$^{23}$ W Hz$^{-1}$ & Condon (1987)\\  
$S_{100}$& 6.4 Jy &  IRAS Faint Source Cat. \\
$S_{60}$& 21.6 Jy &  IRAS Faint Source Cat.  \\
$S_{60}/S_{100}$& 0.30   &  \\
$\lfir$$^2$       & 3.9  10$^{37}$ W  &   \\
$B_T^0$       &  11.01  & RC3\\
$\lb$          &     9.5 10$^{37}$ W  \\
$(B-V)_T^0$ & 0.57 & RC3 \\
$M_{\hi}$  &  1.2\, 10$^{11}$  & SHSVV$^1$ \\
$M_{\hii}$ &  4\, 10$^{10}$ & Young et al. (1995)$^3$ \\
$M_{total}$ & 2.3 \,10$^{12}$ & SHSVV$^1$ \\
\end{tabular}

$^1$ adjusted to $H_0=50$ km s$^{-1}$ Mpc$^{-1}$ \newline
$^2$ $\lfir=1.5\, 10^{32} [D/{\rm Mpc})^2
(2.58 (S_{60}/{\rm Jy})+ (S_{100}/{\rm Jy}))]$ \newline
$^3$ assuming a conversion factor of $M_{\hii}/L_{CO}=4.8$ 
\end{table}
In Table 2 observational properties of NGC~1961 are presented. 
Because of  its size, NGC~1961 is luminous at all wavelengths. 
The $(B-V)_T^0$ colour is typical for an Sb galaxy 
(Huchra 1977) as is the ratio $\lfir/\lb=0.3$ (e.g. Lisenfeld et al. 1996b).
This latter value is a measure of the ratio of 
the SFR averaged over the last $10^8$ yr to the SFR
averaged over the last $10^9$ yr, and indicates that
the {\it global} SF has proceeded  constantly during the past $10^9$ yr. 

The ratio of the far-infrared (FIR) to radio luminosity, 
$\log(\lfir/\pnu)=14.44$, is a factor of 3-5 lower than for 
spiral galaxies. The average values derived
for galaxy samples range from $14.9$ (bright galaxies, 
Condon Anderson \& Helou 1991) to $15.1$ (normal 
spiral galaxies, e.g. Condon et al. 1991, Lisenfeld et al. 1996a) with a 
typical standard deviation of 0.2 in the logarithm.
This result is most likely due to an unusually high radio luminosity; 
the FIR luminosity does not appear unusually low
compared to its blue luminosity ($\lfir/\lb=0.3$), or the
inferred H$_{2}$ mass  ($\lfir/M_{H_{2}}=2 \ls \ms^{-1}$;  for our
Galaxy $\lfir/M_{H_{2}}=3.7 \ls \ms^{-1}$, Solomon
et al. 1992). 
On the other hand, the radio luminosity is high compared
to luminosities in other bands.  For example the radio to blue luminosity
ratio $\prad/\lb=6.25\,10^{11} {\rm W Hz}^{-1} \ls^{-1}$
is a factor of 7.5 higher than found for a sample of Sb galaxies
(Hummel et al. 1988). 
Furthermore, the radio surface brightness is high as reflected in
the magnetic field strength, $B=15 \mu$G,
estimated using the minimum energy method in which 
$B \propto (\prad/ \rm area)^{2/7}$. 
Hummel et al. (1988) derived for a sample of Sb galaxies
(with the same assumption for the minimum energy calculation)
an average value of 8 $\mu$G.

Deviations from the FIR/radio correlation have been found before
for cluster  (Niklas, Klein \& Wielebinski 1995) and interacting
galaxies (NGC~2276, Hummel \& Beck  1995). Also in these
cases the conclusion by the authors was that the radio
emission is enhanced compared to normal galaxies.
The reasons for this enhancement have been suggested to 
be either an additional production
of CRs by shock acceleration at interstellar shocks produced
by the interaction of the galaxy with the intercluster
medium (V\"olk \& Xu 1993), or  an enhanced magnetic field (Hummel \& Beck
1995).

\subsection{The spectral analysis}

The diffusion scale of the CR electrons in NGC~1961 is 
approximately 3.7~kpc at 1.46 GHz 
(taking $B=15 \mu$G, $\urad=1.5$ eV cm$^{-3}$,
and a typical diffusion coefficient
of $10^{29}$cm$^2$s$^{-1}$).
At the distance of NGC 1961 the 16-arcsec beam corresponds to 6.4 kpc;
to a good approximation, therefore, the observations average
over the synchrotron emission of both newly accelerated and aged 
CR electrons. The spectral variations reflect
variations in the thermal fraction,
differences in the conditions of the ISM
and in the CR injection, but not effects due to propagation of the CR electrons.
 
\subsubsection{Star formation history}

The shape of the non-thermal spectrum reflects temporal variations 
of past CR acceleration and therefore of the star-formation history if
SN remnants are
the most important source for the acceleration of CR electrons.
The steep synchrotron spectrum at the centre 
and in the south-eastern arm (positions 3 and 4 in Fig. 7)
indicate aged CR electrons;  the spectra cannot be
explained by continous CR acceleration since this would require an 
injection spectral index of $\gamma=3.4$.
In the region between the centre and the south-eastern arm the synchrotron 
spectrum is much flatter, indicating recent electron acceleration.
The electron age distribution derived from the spectral fitting analysis
gives a time since acceleration for the CR electron population at the
maximum of the radio emission and in the south-eastern
arm (positions 3 and 4) of $\sim 10^7$ yr,  whereas the electrons 
between these sites are considerably younger ($ 3\, \times 10^6$ yr). 

This leads to the following picture of the star formation history.
The end of CR injection $10^7$ yr ago
indicates that the SN activity ceased at that time.
The life-time of SN
progenitor stars is at maximum $8\,\times 10^7$ years (corresponding to
a stellar mass of 5 $\ms$), which suggests an 
epoch of intense  SF  which ended  $\la 9\, \times 10^7$ yr  ago.
The beginning of the star formation epoch cannot be constrained
by our data. 
The recent acceleration of CR electrons deduced from the
flat spectral index of the extended radio emission means that
the CR injection and thus SN activity at these sites has 
recently started. The spectra are too flat to be explained by
continous CR acceleration because in this case the contribution from 
aged CR electrons to the radio emission would steepen the spectrum,
and approximating this situation to that of a steady-state we would
expect a synchrotron spectral index of
$\anth=\gamma/2$. Taking  $\gamma = 2$, the expected synchrotron 
spectral index on the
basis of a steady state model is much steeper than the observed one,
which is,
attributing as an upper limit all of the observed flux density  at 
15~GHz to the thermal emission, $\alpha_{nth,1.46-4.92GHz}=0.4\pm0.2$.
These results are most consistent with a recent burst of star formation
$2 - 3 \times 10^{7}$yr ago followed by a peak SN rate about
$5 \times 10^{6}$yr ago.

The time-scales derived from this analysis imply that a causal
connection between the star formation epochs is possible,
in the sense that the earlier star formation period in the arm 
may have triggered an episode of star formation in the inter-arm region.
The width of the south-eastern arm is about 10 kpc
and taking a typical sound speed to be 10 km/s a SN shock 
would take approximately $10^8$ yr to cross the arm which is consistent
with the time-scale inferred for the star formation history.

SHSVV suggested that  an encounter of NGC 1961 
with an intergalactic  cloud took place less than $5\,10^8$ yr ago
(the rotation time-scale), and triggered
an episode of intense star formation in the south-eastern arm.
They presented an optical colour image showing that the south-eastern arm
has indeed a blue colour,
indicating recent star-formation. This is consistent with the 
star-formation history that we infer.


\subsubsection{Variations of the magnetic field}

In the above analysis we have assumed that the magnetic field strength is 
constant over the galactic disk; this is very likely not the case.
Compression of the ISM and/or the presence of shocks throughout the
disc will lead to enhanced magnetic field strengths and hence increased 
radio synchrotron luminosity relative to FIR.
Such processes were inferred by SHSVV from the morphology of the \hi
distribution.
However, an enhanced magnetic field strength in the south-eastern arm 
alone cannot explain the large radio-to-FIR ratio since the 
radio flux density  at 1.46 GHz of the south-eastern 
arm contributes only about 
15~$\%$ to the total flux density of the galaxy.

If the magnetic field varies on scales shorter than that
of the CR electron propagation it has an effect not only on the
local synchrotron emission but also on the integrated spectral index.
In particular, if the magnetic field strength
decreases (increases) away from the source of CR electrons, the spectrum  
can become 
significantly flatter (steeper). The problem can be
solved in a one-dimensional model in which the magnetic field strength and the 
energy density of the radiation field change with increasing height
over the galactic disk, z. 
Assuming that the magnetic field strength varies as
$B(z)\propto z^{-\sigma}$, the energy losses as
$(\urad+\ub)(z)\propto z^{-\rho}$ and the energy dependence
of the diffusion coefficient is $D(E)\propto E^{\mu}$,
then the resulting spectral index is, in the steady state,
\begin{equation}
\alpha={\gamma\over 2}+{1-\mu\over 2} {\rho - {\sigma\over 2}(2+\gamma)
\over 2-\rho+{\sigma\over 2} (1-\mu)}
\end{equation}
(Berezinsky et al. 1990). 
Taking $\mu=0$, we see that in the case $\rho=0$ 
(i.e. spatially constant inverse Compton losses dominate the energy
losses)
the difference of the spectral index with respect
to the case of a constant magnetic field is
\begin{equation}
\triangle\alpha=\alpha-{\gamma\over 2}=-{1\over 2} {{\sigma\over 2}(2+\gamma)
\over 2+{\sigma\over 2}}.
\end{equation}
For an increasing magnetic field with $\sigma =-1$ this yields
$\triangle\alpha=0.4$ and for 
$\sigma=-2$ $\triangle\alpha=0.66$. Thus a significant steepening 
of the spectrum is possible. If however
synchrotron losses ($\propto B^2$) dominate 
then, since $U_{rad} + U_{B} \approx U_{B} \propto B^{2}$,
we set  $\rho=2 \sigma$ and get:
\begin{equation}
\triangle\alpha={\sigma(1 - {1\over 4}(2+\gamma))
\over 1-\sigma+{\sigma\over 4}}
\end{equation}
In this case only a very weak steepening of the spectrum is
possible; for $\gamma =2$  $\triangle\alpha$ is zero for all values
of $\rho$ and $\sigma$.

For NGC~1961 the high radio surface brightness suggests that synchrotron losses
dominate.  In this case, the above analysis shows that 
under steady-state conditions
variations in the field strength do not lead to significant variation in
the spectral index.  Steep spectra in the south-east arm could,
however,  result from
a non-steady-state scenario in which enhancement of the field was
followed by a burst of star formation with increased synchrotron loss
rate -- it is not possible to modify the spectrum significantly without
incorporating into a model variation in the star-formation rate.

\subsection{A dynamical history for NGC 1961}
 
SHSVV suggest that a collision
with an intergalactic cloud has resulted in the unusual
\hi distribution in NGC~1961.
Such an interaction with the IGM can explain
in addition to the optical and \hi morphology
many features of the radio emission provided the interaction
triggered an episode of intense SF in the south-eastern arm.
The main difficulty with this model is that the massive intergalactic
cloud required has not been
observed either as ionized (Pence \& Rots 1997) or in neutral 
(SHSVV) gas.

An alternative view is that the disturbed appearance of NGC 1961
could be the result of a merger.
SHSVV discarded this possibility since
they found no evidence for it either in the optical (i.e. there
is no second nucleus) or in the \hi distribution. However, our
high resolution radio images reveal the presence of a second nucleus
in the centre of the galaxy consistent with an advanced merger; the disturbed
optical and \hi appearance is then naturally explained as a merger remnant.
Near-infrared observations will provide an important indication as
to whether a highly obscured optical nucleus is present.

\section{Conclusions}

We have presented radio data at 4 frequencies of the 
supermassive spiral galaxy NGC 1961. Using these data
we fitted a combined aged-synchrotron plus thermal emission
spectrum to each point in the galaxy. 
The most recent star formation, that is traced by the thermal radio emission,
is mainly taking place in the centre and the south-eastern arm of the
galaxy, whereas the underlying extended radio emission is predominantly
nonthermal. 

The  spectral index distribution between 1.46 and 4.92~GHz
over the disk of the galaxy
shows a peculiar behaviour. At the maxima of the radio emission 
the nonthermal spectral index is very steep whereas the extended 
radio emission has a much flatter spectrum. 
We discuss various possibilities to explain this peculiar behaviour.
The most likely explanation is that the steep spectrum is
due to the end of the CR injection about $10^7$ years ago. 
Assuming that CRs are accelerated mainly in SNRs this means that
a phase  of intense SF has ceased  $\la 10^8$ yr ago.
The flat spectrum of the extended emission is a sign of  recent
CR acceleration in these regions. This could be due to CR acceleration 
in the shocks of SNRs of a  recently formed stellar
population. The radio-to-FIR ratio
exceeds the average value of spiral galaxies by
a factor of 3-4. An overall increase of the magnetic field
with respect to the radiation field could account for this high
FIR/radio ratio. 
Spatial variations of the magnetic field, which are likely to be present
if the magnetic field is frozen into the gas, are however 
unable to give the observed distribution of the spectral index.

The discovery of a second radio nucleus leads us to suggest that
NGC~1961 is most likely the result of a merger rather than
collision with a massive intergalactic gas cloud.

\section*{ACKNOWLEGDEMENTS}

This research has made use of the NASA/IPAC extragalactic database (NED)
which is operated by the Jet Propulsion Laboratory, Caltech, under
contract with the National Aeronautics and Space Administration.
UL gratefully acknowledges the receipt of a grant of the
Deutsche Forschungsgemeinschaft (DFG) and by the Comisi\'on 
Interministerial de Ciencia y Technolog\'\i a (Spain). 

{}


\begin{thebibliography}{}
\bibitem[]{got}
Becker R.H., White R.L., Edwards A.L., 1991, \apjs 75, 1
\bibitem[]{got}
Berezinsky V.S., Bulanow S.V., GinzburgV.L., Dogiel V.A., Ptsuskin V.S.,
1990, Astrophysics of Cosmic Rays, North Holland, p. 189
\bibitem[]{got}
Condon J.J., 1983, \apjs 53, 459
\bibitem[]{got}
Condon J.J., 1987, \apjs 65, 485
\bibitem[]{got}
Condon J.J., Anderson M.L., Helou G., 1991, \apj 376, 95
\bibitem[]{got}
de Vaucouleurs G., de Vaucouleurs A., Corwin H., et al., 1991,
{\it Third Reference Catalog of Galaxies}, Springer-Verlag, New York (RC3)
\bibitem[]{got}
Gioia I.M., Gregorini L., Klein U., 1982, \aajou 116, 164
\bibitem[]{got}
Gottesmann S.T., Hunter Jr. J.H., Shostak G.S., 1983, \mnras 202, 21p
\bibitem[]{got}
Howarth N.A., 1990, Ph.D. thesis, University of Cambridge
\bibitem[]{got}
Huchra J.P., 1977, \apjs 35, 171
\bibitem[]{got}
Hummel E., Davies R.D., Wolstencroft R.D., van der Hulst J.M., Pedlar A.,
1988, \aajou 199, 91
\bibitem[]{got}
Hummel E., Beck R., 1995, \aajou 303, 691
\bibitem[]{got}
IRAS Faint Source Catalog, 1990, Version 2.0,
Morshir M., Kopon G., Conrow T., et al., Infrared Processing and
Analysis Center
\bibitem[]{got}
Israel F.P., Mahoney M.J., 1990, \apj 352, 30
\bibitem[]{got}
Jaffe W. J., Perola G.C., 1974, \aajou 26, 423
\bibitem[]{got}
Lisenfeld U., V\"olk H.J., Xu C., 1996a, \aajou 306, 677
\bibitem[]{got}
Lisenfeld U., V\"olk H.J., Xu C., 1996b, \aajou 314, 745
\bibitem[]{got}
Niklas S., Klein U., Wielebinski R., 1995, \aajou 293, 56
\bibitem[] {got}
Niklas S., Klein U., Braine J., Wielebinski R., 1995, \aas 114, 21
\bibitem[]{got}
Pence W., Rots A., 1997, ApJ, 478, 107
\bibitem[]{got}
Romanishin W., 1983, \mnras 204, 909
\bibitem[]{got}
Shostak G.S., Hummel E., Shaver P.A., van der Hulst J.M., van der Kruit 
P.C., 1982, \aajou 115, 293  (SHSVV)
\bibitem[]{got}
V\"olk H.J., Xu C., 1993, Infrared Phys.Technol. 35, 527
\bibitem[]{got}
White R.L., Becker R.H., 1992, \apjs 79, 331
\bibitem[]{got}
Young J., Xie S., Tacconi L., et al., 1995, \apjs 98, 219
\end{thebibliography}
\end{document}